\documentclass[11pt]{amsart}

\usepackage{amsmath, amsthm, amssymb, amsfonts, enumerate}
\usepackage[colorlinks=true,linkcolor=blue,urlcolor=blue]{hyperref}
\usepackage{dsfont}
\usepackage{color}
\usepackage{geometry}
\usepackage{todonotes}
\usepackage{graphicx}
\usepackage{subcaption} 
\usepackage{caption}
\usepackage{multirow}
\usepackage{booktabs}
\usepackage{float}
\usepackage{mathtools}
\mathtoolsset{showonlyrefs} 
\usepackage{accents}

\def \cC{\mathcal{C}}

\def \cF{\mathcal{F}}
\def \cG{\mathcal{G}}
\def \cH{\mathcal{H}}

\def \cL{\mathcal{L}}

\def \cS{\mathcal{S}}
\def \cT{\mathcal{T}}

\def \bF{\mathbb{F}}
\def \bG{\mathbb{G}}

\def \P{\mathsf P}

\def \E{\mathsf E}

\def \R{\mathbb{R}}

\def \ud{\mathrm{d}}
\def \e{\mathrm{e}}
\newcommand{\operL}{\ensuremath{\mathbb{L}}}

\newcommand{\ind}{\mathbf{1}}

\newtheorem{theorem}{Theorem}[section]

\newtheorem{proposition}[theorem]{Proposition}

\newtheorem{assumption}[theorem]{Assumption}

\title[Optimal Annuitization Time under a mortality shock]{Optimal Annuitization Time under a mortality shock}
\author[Buttarazzi]{Matteo Buttarazzi}
\address{M.~Buttarazzi: School of Management and Economics, Dept. ESOMAS, University of Torino, Corso
Unione Sovietica, 218 Bis, 10134, Torino, Italy.}
\email{\href{mailto:matteo.buttarazzi@unito.it}{matteo.buttarazzi@unito.it}}
\date{\today}

\numberwithin{equation}{section}

\begin{document}

\begin{abstract}
In this paper, we derive explicit closed-form solutions for the value function and the associated optimal stopping boundaries in an optimal annuitization problem under a mortality shock. We consider an individual whose retirement wealth is invested in a financial fund following the dynamics of a geometric Brownian motion and has the option at any time to irreversibly convert their wealth into a life annuity. The individual faces a sudden, permanent health deterioration occurring at a random, exponentially distributed time, and the annuitization decision is modelled as an optimal stopping problem across two health states. Building on the framework of \cite{buttarazzi2025optimal}, our analytical expressions characterise both the value function and the optimal timing of annuitization. The results provide clear economic intuition: the optimal strategy is governed by the critical interplay between the relative attractiveness of the annuity (money's worth), the financial returns from the investment fund, and bequest motives across different health states. A numerical analysis compares the optimal annuitization strategy of an individual facing a health shock against a benchmark case with constant mortality, highlighting how the likelihood and severity of a health shock significantly alter optimal annuitization behaviour.
\end{abstract}

\maketitle

\section{Introduction}
Annuities are insurance products that convert a lump sum or series of premiums into a guaranteed lifelong income stream. By providing income for life, they protect against longevity risk -- that is, the risk that an individual will exhaust their financial resources before death. 
However, annuitisation is usually an irreversible transaction: once funds are converted into an annuity, the associated capital is typically inaccessible, limiting flexibility in the face of unexpected needs or changing market conditions. Some contracts may permit early withdrawals, but these often come with substantial penalties, see \cite{de2024variable} or \cite{landriault2021high}.
A natural alternative is a self-managed investment strategy, where individuals rely on a portfolio of assets to fund retirement. This \textit{do-it-yourself} approach offers flexibility and control, as the portfolio can be adjusted to personal needs or economic conditions. Yet it exposes retirees fully to both longevity risk and market risk. This creates a trade-off between guaranteed longevity protection on the one hand, and financial flexibility and the potential for higher returns on the other. Bequest motives further shape annuitization decisions. Because annuities typically provide no residual value at death, they are less appealing to individuals with a strong desire to leave a legacy. A substantial body of literature, including work by \cite{bernheim1991strong}, \cite{brown2001private}, \cite{davidoff2005annuities}, and \cite{lockwood2012bequest}, consistently shows that bequest motives substantially reduce the attractiveness of annuities. 
The pricing and evaluation of annuities hinge on \textit{force of mortality}. Consistent with previous studies on optimal annuitization, such as those by \cite{milevsky2007annuitization}, \cite{hainaut2014optimal}, \cite{de2019free} or \cite{buttarazzi2025optimal}, a critical asymmetry arises in this context. Insurers employ an objective force derived from population-level mortality data to price annuities. By contrast, individuals use a subjective force, based on personal health and beliefs, to assess the annuity's value and their expected future cashflows. This asymmetry generates a complex mechanism where the decision to annuitize hinges not only on financial considerations and bequest motives but also on life expectancy and the perceived fairness of the annuity price.

Since Yaari’s seminal work \cite{yaari1965uncertain}, optimal annuitization has been widely studied, see \cite{stabile2006optimal}, \cite{hainaut2006life}, \cite{milevsky2006optimal}, \cite{milevsky2007annuitization}, \cite{hainaut2014optimal}, \cite{de2019free} among others. Hainaut and Deelstra \cite{hainaut2014optimal} analyse an all-or-nothing setting where an individual can irreversibly convert their entire wealth into an immediate annuity. Their wealth is initially invested in a financial fund modelled as a jump-diffusion process, and mortality is introduced via time-dependent objective and subjective forces. The problem is formulated as an optimal-stopping problem and solved numerically. De Angelis and Stabile \cite{de2019free} examine the same setup with a fund following geometric Brownian motion and provide a rigorous probabilistic characterisation of the optimal stopping boundary together with numerical sensitivity analysis.
Building on \cite{hainaut2014optimal} and \cite{de2019free}, who use a deterministic mortality force, \cite{buttarazzi2025optimal} extends their framework by modelling the subjective mortality force as a piecewise deterministic Markov process (introduced by M.H.A. Davis in \cite{davis2018markov}). The mortality force takes constant random values between random jump times, allowing the model to capture a range of health dynamics, from sudden shocks (large, infrequent jumps) to a gradual, random worsening of life conditions (frequent small jumps). 
Within this framework, an individual invests in a fund (modelled as geometric Brownian motion) and must decide when to irreversibly convert all wealth into an annuity to maximise the expected discounted value of dividends, annuity payments, and bequest. They incorporate a bequest motive via a linear payoff for heirs in case of death before annuitization, modulated by a parameter $\nu \in [0,1]$ measuring the intensity of bequest motives. 
Through a careful probabilistic analysis, \cite{buttarazzi2025optimal} provides a complete characterisation of the solution. Specifically, depending on health state, optimal annuitization may occur in any of the following, mutually exclusive, situations: (i) when the retirement wealth exceeds an upper threshold, (ii) when it falls below a lower threshold, (iii) when it exits a given interval of value, (iv) never, no matter the retirement wealth or (v) immediately, no matter the retirement wealth.

This paper applies the general theoretical framework developed in \cite{buttarazzi2025optimal} to a specific and more tractable case. We focus on an individual whose mortality is initially in a low state ($l$), reflecting good health. At a random time, however, a sudden and irreversible health shock may occur, moving the individual into a high-mortality state ($h$) that represents a permanent deterioration in health. 
This setting allows us to derive explicit analytical expressions for the value function and the optimal stopping boundary, offering clear economic interpretation and intuition that complements the broader theoretical results of the general model. 
The problem is first formulated as a two-dimensional optimal stopping problem, with the state variables given by the individual’s wealth and health status. Applying the methodology of \cite{buttarazzi2025optimal}, the problem is then transformed into two nested one-dimensional optimal stopping problems, one for each health state.  The problem is solved backwards, beginning with the high-mortality state ($h$). Since no further health shocks are possible in this state, the problem reduces to an optimal stopping problem with a constant mortality force, which has been solved in closed form by \cite{buttarazzi2024market} (cf. also \cite{mythesis}). Thus, we leverage their results to fully characterise the value function and optimal annuitization strategies in state ($l$). Indeed, the solution for the high-mortality state is a key input for the problem in the low-mortality state. Here, the individual must weigh the value of annuitizing immediately against the value of waiting, which incorporates the option to annuitize later under either the current favourable health conditions or the less favourable ones following a potential health shock. 

The remainder of the paper is structured as follows. Section \ref{problemformulation} presents the model formulation, detailing the financial and mortality dynamics. The main contribution is presented in Section \ref{Sec:EXSOL}, where we derive explicit analytical solutions for the value function and optimal annuitization boundaries for an individual facing a health shock. In Section \ref{numerics}, we conduct a numerical analysis comparing the optimal strategy of this individual with that of a benchmark individual with a constant mortality force $\mu=\mu_l$. We also perform a sensitivity analysis to investigate how jump-related parameters (the severity of the health shock and its intensity) influence the timing of the annuitization decision. Our results offer economic and financial insights that would be difficult to extract from the general solution of the more complex model of \cite{buttarazzi2025optimal}.
Finally, we recall in Appendix \ref{APP:VFconstantexplicit} the explicit solution for the value function and the associated stopping and continuation regions under a constant mortality force $\mu$, as derived in \cite{mythesis}.

\section{Problem Formulation} \label{problemformulation}

\subsection{Financial and actuarial aspects} Consider a probability space $(\Omega,\cG,\P)$ equipped with a Brownian motion $(B_t)_{t\ge 0}$. 
Let $\bF^B\coloneqq(\cF^B_t)_{t\geq 0}$ denote the right-continuous completion of the Brownian filtration with the $\P$-null sets and with
$\cF^B_\infty\coloneqq\vee_{t\geq 0} \cF^B_t$. 
The individual's retirement wealth is invested in a financial fund whose dynamics follow 
\begin{equation}
\left\{ \begin{array}{ll} \ud X_t = (\theta-\alpha)X_t \ud t+\sigma X_t \ud B_t, \ \ \ \ \ \ \ t>0,\\
X_0 = x > 0, \end{array} \right.
\label{Eq:X}
\end{equation}
where $\theta>0$ is the average continuous return of the financial investment, $\alpha\geq 0$ is the constant dividend rate and $\sigma>0$ is the volatility coefficient.

The individual is initially in a low-mortality state ($l$), associated with a mortality force $\mu_l$. A sudden health event may occur at a random time $\xi$, modelled as an exponential random variable with intensity $\lambda_l > 0$. When the shock occurs, the individual’s health state switches irreversibly to a high-mortality state ($h$), and the mortality force jumps to $\mu_h \coloneqq \mu_l + \Delta$, with $\Delta \ge 0$ capturing the severity of the health deterioration. The high-mortality state is absorbing, meaning that no further shocks can occur, we set $\lambda_h=0$.
Letting the health-state process be
$H_t = l \mathds{1}_{\{t < \xi\}}+ h \mathds{1}_{\{t \ge \xi\}}$, the \emph{subjective mortality force process} is then given by
\begin{align} \label{defmu_t}
\mu_t = \mu_l\,\mathbf{1}_{\{H_t=l\}} + \mu_h\,\mathbf{1}_{\{H_t=h\}}, 
\qquad t \ge 0.
\end{align}
For notational convenience, any quantity indexed by $l$ or $h$ is associated with the corresponding health state throughout the paper.

To track the occurrence of the health shock over time, we introduce the filtration $\mathbb{H} \coloneqq (\mathcal{H}_t)_{t \geq 0}$, with $\mathcal{H}_t \coloneqq \sigma(H_s : s \leq t)$ for $t \geq 0$, and $\mathcal{H}_\infty \coloneqq \vee_{t \geq 0} \mathcal{H}_t$. Next, we introduce the assumption of independence between financial and mortality risks.
\begin{assumption} \label{Ass:indipendence}
The $\sigma$-algebras $\cH_\infty$ and $\cF^B_\infty$ are independent. 
\end{assumption}

Let $\bF\coloneqq(\cF_t)_{t\geq0}$ with
$\cF_t=\cF^B_t\vee\cH_t$ for $t\ge 0$, and $\cF_\infty=\vee_{t\geq0}\cF_t$. 
The enlarged filtration $\bF$ contains all the information about both the financial market and the individual's health status. 
Assumption \eqref{Ass:indipendence} ensures that the filtration $\bF$ is right-continuous, see \cite[Ch.\ 1.1.4, Prop.\ 1.12, p.\ 6]{aksamit2017enlargement}.
Then, we define the individual's time of death $\tau_d$ following the canonical construction for a doubly stochastic random time (see, e.g., \cite[Ch.\ 2.3.1]{aksamit2017enlargement}). Letting $\Theta$ be a random variable independent of $\cF_\infty$ and exponentially distributed with parameter one, the death time is defined as \begin{equation} 
\tau_d\coloneqq\inf \Big\{t\geq 0: \int_0^t \mu_s\ud s \geq \Theta\Big\}.
\end{equation}

At time $t\ge0$, the individual uses their \emph{subjective} mortality force process $\mu_t$ to compute their self-assessed life expectancy $_z p_{t}$ defined for $t,z\ge 0$ as follows
\begin{align}\label{Def:Subj_prob}
\begin{aligned}
_z p_{t}&\coloneqq\P(\tau_d>t+z\mid \cF_{t+z},\tau_d>t)\\
&\coloneqq\frac{\P (\tau_d>t+z,\tau_d>t| \cF_{t+z})}{\P(\tau_d>t\mid \cF_{t+z})}=\exp \Big( -\int_{t}^{t+z}\mu_s\ud s\Big),
\end{aligned}
\end{align}
where the last equality follows from the properties of the double stochastic Poisson setting (See \cite[Ch.\ 2.3.2, Lemma 2.25, p. 42]{aksamit2017enlargement}).

The insurer relies on the \textit{objective} mortality force $\hat{\mu}>0$, which is public information derived from demographic analysis of the population. Thus, it is assumed that the insurer has no knowledge of the individual’s personal health condition and instead relies on population-level data.
According to standard actuarial theory, the value at time $t\ge0$ of a life annuity paying at a rate of one monetary unit per year is given by:
\begin{equation} \label{ahatintro}
\hat{a}_{t}\coloneqq \int_{0}^{\infty} \e^{-(\hat{\rho}+\hat{\mu}) s} ds=\frac{1}{\hat{\rho}+\hat{\mu}},
\end{equation}
where $\hat{\rho}>0$ is the interest rate guaranteed by the insurer. On the other hand, the individual evaluates the annuity's attractiveness by using the coefficient 
\begin{align} \label{aintro}
    a_{t}\coloneqq \E \Big[ \int_{0}^{\infty} \e^{-\rho u} {}_u p_{t}\ \ud u \mid \cF_t \Big].
\end{align}
The ratio $a_{t}/\hat{a}_t$, known in actuarial science as "money's worth", gives at any time the ratio between the individual's valuation of the unitary annuity and its market price. It is directly related to the gap between subjective and objective mortality force. For individuals who are on average healthier (or less healthy) than the reference population, this ratio is greater (or lower) than 1, and the annuity is underpriced (or overpriced). When $a_{t}/\hat{a}_t=1$, the annuity is {\em actuarially fair}.
In the next proposition, we derive an explicit expression for the money’s worth. The proof is provided in the supplementary material.

\begin{proposition} \label{Prop:aoverahat}
For $t \ge 0$, the money's worth $a_{t}/\hat{a}_t$ can be rewritten as
\begin{align}
\begin{aligned} \label{defAdelta0}
    \frac{a_{t}}{\hat{a}_{t}} = \mathds{1}_{\{t < \xi\}}\, \delta_l + \mathds{1}_{\{t \geq \xi\}}\, \delta_h, 
\end{aligned}
\end{align}
where for $i\in\{l,h\}$
\begin{align} \label{defdelta0}
    \delta_i\coloneqq \frac{(\rho + \lambda_i + \mu_h)(\hat{\rho}+\hat{\mu})}{(\rho + \lambda_i + \mu_i) (\rho+\mu_h)}.
\end{align}
\end{proposition}

\subsection{Optimization Problem}
The individual invests in the financial market and collects dividends at a constant rate $\alpha$. At a random time $\tau:\Omega\to[0,\infty]$, they may choose to convert their entire wealth into an immediate lifetime annuity. 
After annuitization, i.e. for $t>\tau$, the individual receives the annuity payments at a constant annual rate 
\begin{align}\begin{aligned} \label{annuitypayments}
P_{\tau}\coloneqq \frac{X_\tau-K}{\hat{a}_{\tau}},
\end{aligned}\end{align}
where $K$ represents either a fixed acquisition tax ($K>0$) or an incentive ($K<0$).
Should the individual die before the annuitization time, i.e. on the event $\{\tau\ge \tau_d\}$, they leave a bequest equal to their wealth at the time of death.

The individual weighs the cash flows using a linear utility. Since we are interested in the impact of the strength of bequest motives on the individual's annuitization decision, with no loss of mathematical generality, we assume that the linear utility of the dividend and annuity payments has a unitary slope; instead, the utility from bequest is modulated by a parameter $\nu\in[0,1]$, measuring the strength of bequest motives. 

Let the filtration $\bG$ be such that each $\sigma$-algebra $\cG_t$ is defined as $\cG_t\coloneqq\cap_{s>t}(\cF_s \vee \sigma(\tau_d\wedge s))$ and denote by $\cT(\bG)$ the set of $\bG$-stopping times.
The individual aims to maximise their expected payoff by optimally choosing the time of the annuity purchase.
The optimisation problem reads
\begin{align}  \label{vf0}
    \begin{aligned}
    V_0=\sup_{\tau \in \cT(\bG)} \E \Big[\int_{0}^{\tau_d \wedge \tau }\e^{-\rho t}\alpha X_t \ud t  +\mathds{1}_{\{\tau_d \leq \tau \}}\e^{-\rho \tau_d} \nu X_{\tau_d} +P_{\tau}\int_{\tau_d\wedge \tau}^{\tau_d}\e^{-\rho t}\ud t \Big].
    \end{aligned}
\end{align}

The following assumption ensures that the above problem is well-posed (cf. \cite[Prop. 5.1]{buttarazzi2025optimal}).
\begin{assumption}
    \label{mainass} $\theta-\alpha-\rho-\mu_l<0$. 
\end{assumption}

The problem \eqref{vf0} can be cast in a Markovian framework, see \cite[Prop. 2.5]{buttarazzi2025optimal}. In particular $V_0=V(x,\mu_l)$ for 
\begin{align}
    \begin{aligned}\label{VF:recursive}
        V(x,\mu_l)&= \!\!\sup_{\tau \in\cT(\bF^B)} \E_{x,\mu_l} \Big[ \e^{-r_l \tau}  \delta_l (X_\tau\!-\!K)+\int_0^\tau\!\! \e^{-r_l t} \big[( \alpha\!+\!\nu \mu_l ) X_t \!+\!\lambda_l V(X_t,\mu_h) \big]\ud t  \Big],
\end{aligned}\end{align}
where 
\begin{align}
\begin{aligned}\label{VF:constant}
        V(x,\mu_h)&= \!\!\sup_{\tau \in\cT(\bF^B)} \E_{x,\mu_h} \Big[ \e^{-r_h \tau}  \delta_h (X_\tau\!-\!K)+\int_0^\tau\!\! \e^{-r_h t} ( \alpha\!+\!\nu \mu_h ) X_t \ud t  \Big],
\end{aligned}\end{align}
with $r_i\coloneqq \rho+\mu_i+\lambda_i$ and $\P_{x,\mu_i}(\cdot)=\P(\cdot|X_0=x, \mu_0=\mu_i)$ for $i=\{l,h\}$. 

The reformulation above reduces the problem to an optimal stopping problem on the 1-dimensional diffusion $(X_t)_{t\ge 0}$ and $V(x,\mu_i)$ depends explicitly on the mortality state $\mu_i$ only in a parametric way. 
Moreover, the regularity established in \cite[Thm 2.6]{buttarazzi2025optimal} allows $V(\cdot, \mu_i)$ to be extended to $[0,\infty)$, so that the continuation and stopping regions can be defined as \begin{align} 
&\mathcal{S}_{i}\coloneqq\{x\in [0,\infty): V(x,\mu_i)= \delta_i (x-K)\}, \qquad \mathcal{C}_{i}\coloneqq[0,\infty)\setminus \mathcal{S}_{i}
\end{align}
for $i=\{l,h\}$.
The optimal stopping time is the first entry time of the process $(X_t)_{t\geq0}$ in the stopping region $\cS_i$, i.e. $\tau^*_i\coloneqq \inf \{t\ge 0: X_t \in \cS_i\}$.

The aim of this paper is to derive an explicit characterisation of the value function $V(x,\mu_l)$ and the associated continuation and stopping regions. To this end, we introduce the infinitesimal generator $\operL$ of the process $(X_t)_{t\geq 0}$
\begin{align}\label{L}
(\cL u)(x):=\tfrac{1}{2} \ \sigma^2 x^2 u_{xx}(x)+(\theta-\alpha)x u_{x}(x), \quad \text{for any} \ u\in C^2(\R),
\end{align}
where $u_x=\partial u/\partial x$ and $u_{xx}=\partial^2 u/\partial x^2$.
The explicit solutions are obtained by solving the associated free-boundary problem:
\begin{align}\label{eq:ODE}
\begin{aligned}
\operL V(x,\mu_l) -r_l V(x,\mu_l)=-(\alpha+\nu\mu_l)x-\lambda_l V(x,\mu_h), \quad x\in\cC_l,
\end{aligned}
\end{align}
(cf. \cite[Thm 2.6]{buttarazzi2025optimal}).
It is useful to recall (cf. \cite[Ch.\ II.1, pp.\ 18-19]{borodin2015handbook}) that the fundamental solutions to $\operL u = r_i u$ are given for $i\in\{l,h\}$ by $\psi(x,\mu_i) \coloneqq x^{\gamma^+_i}$ and $\phi(x,\mu_i) \coloneqq x^{\gamma^-_i}$, where 
\begin{equation} \label{gammapm0}
\gamma^{\pm}_i\coloneqq \frac{1}{2} - \frac{\theta - \alpha}{\sigma^2} \pm \sqrt{\left( \frac{1}{2} - \frac{\theta - \alpha}{\sigma^2} \right)^2 + \frac{2 r_i}{\sigma^2}}.
\end{equation}

The value function $V(x,\mu_h)$ corresponds to the case of a constant mortality force $\mu_h$. The characterization of the stopping and continuation regions, $\mathcal{S}_h$ and $\mathcal{C}_h$, is established in \cite{buttarazzi2024market}. Although a closed-form expression for $V(x,\mu_h)$ is not explicitly stated there, it can be directly obtained from the results therein. An alternative derivation is given in \cite[Ch. 3.4]{mythesis}. We recall in Appendix \ref{APP:VFconstantexplicit} the explicit solution for the value function and the associated stopping and continuation regions under a generic constant mortality force $\mu$. This is particularly useful because it enables a clear comparison between the optimal annuitization behaviour of an individual exposed to stochastic mortality risk and that of a benchmark individual facing a constant mortality force $\mu$. Such a comparison, which we carry out in Section \ref{numerics}, highlights the impact of health uncertainty on annuitization choices. In particular, the post-shock value function $V(x,\mu_h)$ required in our analysis follows directly from these generic results by setting $\mu=\mu_h$.

\section{Explicit solutions} \label{Sec:EXSOL}
In this section, we present closed-form expressions for the value function $V(x,\mu_l)$ and the associated continuation and stopping regions. The results are organised into four propositions, covering all parameter configurations of the model. Proofs are deferred to the supplementary material.

For $i\in\{l,h\}$, let 
\begin{align} 
\label{M_i}
    M_i\coloneqq(\delta_i-\beta_i)(\theta-\alpha-r_i ) +\lambda_i \max\{\delta_h,\beta_h\}
\end{align}
where 
\begin{align} \label{def:beta_i}
\beta_i\coloneqq \frac{\alpha+\nu \mu_i}{r_i+\alpha-\theta}.
\end{align}
The coefficient $\beta_i$ reflects the combined effect of dividend payments and bequest motives, scaled by the discounting $r_i$ and the fund’s excess return $(\theta - \alpha)$. The term $\delta_i$ in \ref{defdelta0} corresponds to the individual’s valuation of the annuity relative to its market price (money's worth).
Thus, the term $\delta_l - \beta_l$ measures the attractiveness of annuitization compared to remaining invested in the financial fund in the pre-shock state, while the term $\lambda_l\max\{\delta_h, \beta_h\}$ accounts for the post-shock attractiveness weighted by the shock intensity.
Recalling that $\theta - \alpha - r_l < 0$ (Assumption \ref{mainass}), the sign of $M_l$ provides an indicator of whether the annuity is overall more appealing than staying invested in the fund when the mortality may jump.
Notice that, in the post-shock state, $M_h$ reduces to $(\delta_h-\beta_h)(\theta-\alpha-r_h)$ since $\lambda_h=0$. Assumption \ref{mainass} also implies $\theta-\alpha-r_h<0$ and the sign of $M_h$ coincides with that of $\delta_h-\beta_h$, which determines the structure of the post-shock continuation and stopping regions (cf. Appendix \ref{APP:VFconstantexplicit}).

For the sake of simplifying the derivation of explicit expressions, we assume
\begin{assumption}
    $\Delta \neq \lambda_l$.
\end{assumption}
The exclusion of the case $\Delta=\lambda_l$ is mathematically convenient but theoretically not essential.
In this case the characteristic exponents $\gamma_h^-$ and $\gamma_l^-$ coincide, and the value function $V(x,\mu_l)$ in Propositions \ref{Klowerbeta} and \ref{Prop.2} acquires an additional term of the form $x^{\gamma_l^-}\log x$, which makes the closed-form representation considerably more cumbersome.

\subsection{Case $K<0$}
We focus on the scenario $K<0$, representing an incentive for purchasing the annuity. In the next proposition, we analyse the case $\delta_h \geq \beta_h$, and in Proposition \ref{Klowerbeta} the case $\delta_h < \beta_h$. It is useful to introduce 
\begin{align}
\label{Minftydeltabeta}
M^\delta_l \coloneqq \ind_{\{\delta_h\ge \beta_h\}} M_l, \qquad  M^\beta_l   \coloneqq \ind_{\{\delta_h < \beta_h\}} M_l.
\end{align}
Then, we are now ready to state our main results.

\begin{proposition} \label{propKlowerDelta}
Let $K<0$ and $\delta_h\geq \beta_h$. 
\begin{itemize}
    \item[(i)] If $M^\delta_l\leq 0$, then $\mathcal{C}_l=\varnothing$, $\mathcal{S}_l=[0,\infty)$ and $V(x,\mu_l)=\delta_l (x-K)$.
    \item[(ii)] If $M^\delta_l> 0$, then $\mathcal{C}_l=(x^1_l,\infty)$, $\mathcal{S}_l=[0,x^1_l]$ and
        \begin{align}
            \begin{aligned}
                V(x,\mu_l)=\delta_l (x-K)+\ind_{\{x>x^1_l\}} \Big[ \zeta^1_l x^{\gamma^-_l}  + \frac{ M^\delta_l }{\theta-\alpha-r_l } x - \frac{\hat{\rho} + \hat{\mu}}{r_l } K\Big]
            \end{aligned}
        \end{align}
    with $\gamma^-_l$ given in \eqref{gammapm0}, $M^\delta_l$ in \eqref{Minftydeltabeta}, and
        \begin{align}
            \begin{aligned}
            x^1_l\coloneqq\frac{\gamma^-_l(\hat{\rho} + \hat{\mu})(\theta-\alpha-r_l )K}{(\gamma^-_l-1)r_l M^\delta_l}, 
            \qquad 
            \zeta^1_l\coloneqq-\Big[\frac{(\hat{\rho} + \hat{\mu})K}{(\gamma^-_l-1)r_l}\Big]^{1-\gamma^-_l} \Big[\frac{\gamma^-_l(\theta-\alpha-r_l )}{M^\delta_l}\Big]^{-\gamma^-_l}. 
            \end{aligned}
        \end{align}
\end{itemize}
\end{proposition}

The above proposition describes the scenario where the post-shock condition $\delta_h \geq \beta_h$ holds, meaning that the annuity is sufficiently attractive relative to the investment fund. Combined with the financial incentive to purchase ($K<0$), this leads to an immediate annuitization strategy. Consequently, the stopping region is $\mathcal{S}_h = [0, \infty)$, meaning the individual after the health shock will annuitize at any wealth level (cf. Appendix \ref{APP:VFconstantexplicit}).
The optimal behaviour before the shock depends on the sign of $M^\delta_l$ (equivalently, $M_l$), which determines the relative appeal of the fund versus the annuity. 
When $M^\delta_l \leq 0$, which occurs when $\delta_l$ is sufficiently larger than $\beta_l$, annuitization is more attractive than remaining in the fund, and the individual annuitizes immediately at any wealth level, i.e. $\cS_l = [0,\infty)$.
When $M^\delta_l > 0$, which for instance occurs when $\beta_l \geq \delta_l$, the investment fund offers a valuable alternative. In this case, annuitization is delayed until wealth falls below a threshold $x^1_l$. For wealth levels above this threshold, the individual prefers the financial investment.

\begin{proposition} \label{Klowerbeta}
Let $K<0$ and $\delta_h < \beta_h$.
\begin{itemize}
    \item[(i)] If $M^\beta_l \leq 0$, then $\mathcal{C}_l=\varnothing$, $\mathcal{S}_l=[0,\infty)$ and $V(x,\mu_l)=\delta_l (x-K)$.

    \item[(ii)] If $M^\beta_l>0$, then there is $x^2_l>0$ such that $\mathcal{C}_l=(x^2_l,\infty)$ and $\mathcal{S}_l=[0,x^2_l]$.
    \begin{itemize}
        \item[(ii.1)] If $x^2_l> x^2_h$, then $x^2_l$ solves
        \begin{align} \label{system1xfstar01}
        \frac{(\gamma^-_l-1)M^\beta_l}{\gamma^-_l(\theta-\alpha-r_l)} x + \frac{\lambda_l \zeta^2_h(\gamma^-_l-\gamma^-_h)}{\gamma^-_l(\Delta-\lambda_l)} {x}^{\gamma^-_h} -\delta_lK=0,
        \end{align}  
        where $(\gamma^-_i)_{i=\{l,h\}}$ in \eqref{gammapm0}, $M^\beta_l$ in \eqref{Minftydeltabeta} and
        \begin{align} \label{xdzetad}
        \begin{aligned}
        x^2_h\coloneqq\frac{\delta_h K \gamma^-_h}{(\gamma^-_h-1)(\delta_h-\beta_h)}, \qquad
        \zeta^2_h\coloneqq\Big(\frac{\delta_h K}{\gamma^-_h-1}\Big)^{1-\gamma^-_h}\Big(\frac{\delta_h - \beta_h}{\gamma^-_h}\Big)^{\gamma^-_h}.
        \end{aligned}
        \end{align}
        
        In this case, the value function is
        \begin{align} \begin{aligned} V(x,\mu_l) = \delta_l (x-K)+ \ind_{\left\{ x > x^2_l \right\}}\Big(
        \zeta^2_l x^{\gamma^-_l} - \delta_l K + \frac{M^\beta_l}{\theta - \alpha - r_l} x + \frac{\lambda_l \zeta^2_h}{\Delta - \lambda_l} x^{\gamma^-_h}\Big) \end{aligned}\end{align}
        with 
        \begin{align} \begin{aligned} 
        \zeta^2_l\coloneqq - \frac{M^\beta_l}{\gamma^-_l(\theta - \alpha - r_l)}{(x^2_l)}^{1-\gamma^-_l }-\frac{\gamma^-_h \lambda_l \zeta^2_h}{\gamma^-_l(\Delta-\lambda_l)}{(x^2_l)}^{\gamma^-_h - \gamma^-_l}.
        \end{aligned}\end{align}

        \item[(ii.2)] If $x^2_l \le x^2_h$, then $x^2_l$ solves
        \begin{align}
        \label{system1xfstar11}
        \frac{\gamma^-_l-\gamma^+_l}{\gamma^-_l} \varpi^2_l (x^2_h)^{1-\gamma^+_l} {x}^{\gamma^+_l} +  \frac{\gamma^-_l-1}{\gamma^-_l} \frac{ M^\delta_l }{\theta-\alpha-r_l }x-\frac{\hat{\rho}+\hat{\mu}}{r_l}K=0, 
        \end{align}
        with $M^\delta_l$ defined in \eqref{Minftydeltabeta} and 
        \begin{align} \begin{aligned} \varpi^2_l\coloneqq\Big[\frac{\gamma^-_l-1}{\theta-\alpha-r_l}+\frac{\gamma^-_l(\gamma^-_h-1)}{r_l\gamma^-_h}+\frac{\gamma^-_h-\gamma^-_l}{\gamma^-_h(\Delta-\lambda_l)}\Big]\frac{\lambda_l(\delta_h-\beta_h)}{\gamma^+_l-\gamma^-_l}.\end{aligned}\end{align}
        In this case, the value function is 
        \begin{align} \begin{aligned} 
        V(x,\mu_l) &= \ind_{\{x^2_l<x\le x^2_h\}}\Big[ \varpi^2_l (x^2_h)^{1-\gamma^+_l} x^{\gamma^+_l} + \hat\zeta^2_l  x^{\gamma^-_l} + \frac{M^\delta_l}{\theta - \alpha - r_l} x - \frac{\hat{\rho} + \hat{\mu}}{r_l} K\Big]\\
        &\quad + \ind_{\{x> x^2_h\}} \Big[ \big(\varpi^2_l (x^2_h)^{1-\gamma^-_l} + \pi^2_l (x^2_h)^{-\gamma^-_l}\big) x^{\gamma^-_l} + \hat\zeta^2_l x^{\gamma^-_l} - \delta_l K  \\
        &\quad + \frac{M^\beta_l}{\theta - \alpha - r_l} x + \frac{ \lambda_l \zeta^2_h}{\Delta - \lambda_l} x^{\gamma^-_h}\Big] + \delta_l (x-K)
        \end{aligned}\end{align} 
        with 
        \begin{align} \begin{aligned} \label{pi12}&\hat\zeta^2_l\coloneqq -\frac{\gamma^+_l}{\gamma^-_l} \varpi^2_l (x^2_h)^{1-\gamma^+_l} {(x^2_l)}^{\gamma^+_l-\gamma^-_l} -  \frac{1}{\gamma^-_l} \frac{ M^\delta_l }{\theta-\alpha-r_l} {(x^2_l)}^{1-\gamma^-_l},\\ 
        &\pi^2_l\coloneqq\Big[\frac{1}{r_l} + \frac{\gamma^-_h}{(\theta-\alpha-r_l)(\gamma^-_h-1)}-\frac{1}{(\Delta-\lambda_l)(\gamma^-_h-1)} \Big]\lambda_l \delta_h K.\end{aligned}\end{align}  
        \end{itemize}
    
\end{itemize}
\end{proposition}

The above proposition describes the scenario where, after the health shock, the financial index dominates the annuity's value ($\beta_h > \delta_h$), yet a financial incentive to purchase exists ($K<0$). In the \textit{post}-shock regime, the individual annuitizes only when wealth falls below the threshold $x^2_h$ defined in \eqref{xdzetad}, i.e. $\cS_h=[0,x^2_h]$ (cf.\ Appendix \ref{APP:VFconstantexplicit}).
The optimal strategy \textit{before} the shock is determined by the sign of $M^\beta_l$. If $M^\beta_l\le 0$, immediate annuitization is optimal at any wealth level, that is $\mathcal{S}_l=[0,\infty)$. By contrast, when $M^\beta_l>0$, the individual adopts a threshold strategy, choosing to annuitize only if wealth drops below the threshold $x^2_l$. This pre-shock threshold is defined as the solution to a nonlinear equation, and its relation to the post-shock threshold $x^2_h$ is generally not known in closed form (cf. Proposition \ref{Klowerbeta}--(ii)).
Consequently, the proposition gives rise to two possible configurations depending on the parameters. The first is $\mathcal{S}_l \subseteq \mathcal{S}_h$, which occurs if $x^2_l \le x^2_h$. This means the individual requires a lower wealth level to annuitize before the shock than after it. The second configuration is $\mathcal{S}_l \supset \mathcal{S}_h$, which occurs if $x^2_l > x^2_h$. Here, the individual is willing to annuitize at a higher wealth level pre-shock than post-shock. The specific ordering depends on the model parameters and reflects the individual's complex trade-off between pre-shock investment opportunities and the risk of future health deterioration.

\subsection{Case $K>0$}
In the next proposition, we analyse the case $\delta_h \le \beta_h$, and in Proposition \ref{Prop.2} the case $\delta_h > \beta_h$, when the annuity purchase involves a tax ($K>0$).

\begin{proposition} \label{prop.1}
Let $K>0$ and $\delta_h\leq\beta_h$.
\begin{itemize}
    \item[(i)] If $M^\beta_l<0$, then $\mathcal{C}_l=[0,x^3_l)$, $\mathcal{S}_l=[x^3_l,\infty)$ and 
    \begin{align} \begin{aligned}
    V(x,\mu_l)=\delta_l(x-K)+\ind_{\{x<x^3_l\}} \Big(\zeta^3_l x^{\gamma^+_l}+ \frac{ M^\beta_l }{\theta-\alpha-r_l } x -  \delta_lK\Big)
    \end{aligned}\end{align} 
    with $\gamma^+_l$ given in \eqref{gammapm0}, $M^\beta_l$ in \eqref{Minftydeltabeta}, $\delta_l$ in \eqref{defdelta0} and
    \begin{align}
        \begin{aligned}
        x^3_l\coloneqq\frac{\gamma^+_l(\theta-\alpha-r_l )\delta_l K}{(\gamma^+_l-1) M^\beta_l}, \qquad \zeta^3_l\coloneqq-\Big(\frac{\delta_l K}{\gamma^+_l-1}\Big)^{1-\gamma^+_l} \Big(\frac{\gamma^+_l(\theta-\alpha-r_l )}{M^\beta_l}\Big)^{-\gamma^+_l}.
        \end{aligned}
    \end{align}

    \item[(ii)] If $M^\beta_l\ge 0$, then $\mathcal{C}_l=[0,\infty)$, $\mathcal{S}_l=\varnothing$, and \begin{align}
        V(x,\mu_l)=\Big(\delta_l + \frac{M^\beta_l}{r_l+\alpha-\theta} \Big) x.
    \end{align}
\end{itemize}
\end{proposition}

The above proposition describes the scenario in which the post-shock money's worth of the annuity, $\delta_h$, is dominated by the financial index, $\beta_h$, and the annuity purchase is subject to a fee ($K > 0$). Under these conditions, in the post-shock regime annuitization is never optimal, resulting in an empty stopping region, i.e., $\mathcal{S}_h = \varnothing$ (cf. Appendix \ref{APP:VFconstantexplicit}).
In the pre-shock regime, however, annuitization may still be optimal. Specifically, if $M^\beta_l < 0$, a condition that arises when $\delta_l$ is sufficiently larger than $\beta_l$, the annuity remains more attractive than the financial investment even after accounting for the fee. In this case, the individual will annuitize once wealth exceeds the threshold $x^3_l$. Conversely, if $M^\beta_l \geq 0$, the annuity is not sufficiently attractive to outweigh the financial investment opportunity, and the individual will never annuitize, i.e., $\mathcal{S}_l = \varnothing$.

\begin{proposition} \label{Prop.2}
Let $K>0$ and $\delta_h>\beta_h$.
\begin{itemize}
    \item[(i)] If $M^\delta_l<0$, then there is $x^4_l>0$ such that $\mathcal{C}_l=[0,x^4_l)$ and $\mathcal{S}_l=[x^4_l,\infty)$.
    \begin{itemize}
    \item[(i.1)] If $x^4_l < x^4_h$, then $x^4_l$ solves
    \begin{align} 
    \label{systemforxh01}\\
    \frac{M^\beta_l(\gamma^+_l-1)}{\gamma^+_l(\theta-\alpha-r_l)} x + \frac{ \lambda_l \zeta^4_h(\gamma^+_l-\gamma^+_h)}{\gamma^+_l(\Delta-\lambda_l)} {x}^{\gamma^+_h}-\delta_lK=0,
    \end{align} 
    with 
    \begin{align} 
    \begin{aligned}
    \label{xczetac}
    x^4_h\coloneqq\frac{\delta_h K \gamma^+_h}{(\gamma^+_h-1)(\delta_h-\beta_h)} ,\qquad    \zeta^4_h\coloneqq\Big(\frac{\delta_h K}{\gamma^+_h-1}\Big)^{1-\gamma^+_h}\Big(\frac{\delta_h - \beta_h}{\gamma^+_h}\Big)^{\gamma^+_h}>0.
    \end{aligned}
    \end{align}

    Then, the value function is 
    \begin{align}
        \begin{aligned}
           V(x,\mu_l)=\delta_l(x-K)+\ind_{x<x^4_l} \Big(
           \zeta^4_l x^{\gamma^+_l} -\delta_lK+\frac{M^\beta_l}{\theta-\alpha-r_l} x + \frac{\lambda_l \zeta^4_h}{\Delta-\lambda_l} x^{\gamma^+_h}\Big),
        \end{aligned}
    \end{align}
    with 
    \begin{align} \begin{aligned} 
    \zeta^4_l\coloneqq-\frac{M^\beta_l}{\gamma^+_l(\theta-\alpha-r_l)} {(x^4_l)}^{1-\gamma^+_l} - \frac{ \lambda_l \zeta^4_h \gamma^+_h}{\gamma^+_l(\Delta-\lambda_l)} {(x^4_l)}^{\gamma^+_h-\gamma^+_l}. \end{aligned}\end{align}  

    \item[(i.2)] If $x^4_l\geq x^4_h$, then $x^4_l$ solves
    \begin{align} 
    \label{systemforxh11}\\
    \frac{\gamma^+_l-\gamma^-_l}{\gamma^+_l} \varpi^4_l (x^4_h)^{1-\gamma^-_l} {x}^{\gamma^-_l} +  \frac{\gamma^+_l-1}{\gamma^+_l} \frac{ M^\delta_l }{\theta-\alpha-r_l }x -\frac{\hat{\rho}+\hat{\mu}}{r_l}K=0,
    \end{align} 
    with
    \begin{align} \begin{aligned} \label{pi34}
    \varpi^4_l\coloneqq\Big[\frac{(1-\gamma^+_l)}{\theta-\alpha-r_l}-\frac{\gamma^+_l(\gamma^+_h-1)}{r_l\gamma^+_h}+\frac{\gamma^+_l-\gamma^+_h}{\gamma^+_h(\Delta-\lambda_l)}\Big]\frac{\lambda_l(\delta_h-\beta_h)}{\gamma^+_l-\gamma^-_l}.
    \end{aligned} \end{align} 
    In this case, the value function is
    \begin{align} \begin{aligned} V(x,\mu_l) &=\delta_l(x-K)+
    \ind_{\{x\le x^4_h\}}\Big[\Big(\hat\zeta^4_l  + \varpi^4_l (x^4_h)^{1 - \gamma^+_l} + \pi^4_l (x^4_h)^{-\gamma^+_l}\Big)x^{ \gamma^+_l} \\
    & \hspace{4cm} -\delta_lK+\frac{M^\beta_l}{\theta-\alpha-r_l} x + \frac{ \lambda_l \zeta^4_h}{\Delta-\lambda_l} x^{\gamma^+_h}\Big] \\
    &\quad +\ind_{\{x^4_h<x<x^4_l\}} \Big[\hat\zeta^4_l x^{\gamma^+_l} + \varpi^4_l (x^4_h)^{1-\gamma^-_l} x^{\gamma^-_l} + \frac{ M^\delta_l }{\theta-\alpha-r_l } x -
    \frac{\hat{\rho} + \hat{\mu}}{r_l } K \Big]
    \end{aligned}\end{align} 
    with 
    \begin{align} \begin{aligned} \label{pi34}
    &\hat\zeta^4_l\coloneqq-\frac{\gamma^-_l}{\gamma^+_l} \varpi^4_l (x^4_h)^{1-\gamma^-_l} {(x^4_l)}^{\gamma^-_l-\gamma^+_l} -  \frac{1}{\gamma^+_l} \frac{ M^\delta_l }{\theta-\alpha-r_l } {(x^4_l)}^{1-\gamma^+_l},\\
    &\pi^4_l\coloneqq\Big[\frac{1}{r_l} + \frac{\gamma^+_h}{(\theta-\alpha-r_l)(\gamma^+_h-1)}-\frac{1}{(\Delta-\lambda_l)(\gamma^+_h-1)} \Big]\lambda_l \delta_h K.
    \end{aligned} \end{align} 
    \end{itemize}

    \item[(ii)] If $M^\delta_l\ge 0$, then $\mathcal{C}_l=[0,\infty)$, $\mathcal{S}_l=\varnothing$, and 
    \begin{align}
\begin{aligned} 
    V(x,\mu_l)&=\beta_l x +\lambda_l  x \alpha_1(x) + \lambda_l \zeta^4_h  x^{\gamma^+_h} \alpha_2(x)-\lambda_l \delta_h K \alpha_3(x)
\end{aligned}
\end{align}
with  
\begin{align}
    \begin{aligned}
        \alpha_1 (x) &\coloneqq\frac{\delta_h}{r_l +\alpha-\theta}\ind_{\{x\geq x^4_h\}} + \Big[\frac{\beta_h}{r_l +\alpha-\theta} \\
        &\quad + \frac{(\delta_h-\beta_h)(1-\gamma^+_l)}{(r_l +\alpha-\theta)(\gamma^-_l-\gamma^+_l)} \Big( \Big( \frac{x^4_h}{x} \Big)^{1-\gamma^-_l} -\frac{1-\gamma^-_l}{1-\gamma^+_l} \Big( \frac{x^4_h}{x} \Big)^{1-\gamma^+_l}\Big)\Big]\ind_{\{x<x^4_h\}}\\
        \alpha_2 (x) &\coloneqq\Big[\frac{1}{\lambda_l-\Delta }-\frac{\gamma^+_h- \gamma^+_l }{(\lambda_l-\Delta)(\gamma^-_l-\gamma^+_l)} \Big( \Big( \frac{x^4_h}{x} \Big)^{\gamma^+_h-\gamma^-_l} -\frac{\gamma^+_h-\gamma^-_l}{\gamma^+_h-\gamma^+_l} \Big( \frac{x^4_h}{x} \Big)^{\gamma^+_h- \gamma^+_l }\Big)\Big]\ind_{\{x< x^4_h\}}\\
        \alpha_3(x) &\coloneqq \frac{1}{r_l }\ind_{\{x\geq x^4_h\}} -\frac{\gamma^+_l}{r_l (\gamma^-_l-\gamma^+_l)} \Big[ \Big( \frac{x^4_h}{x} \Big)^{- \gamma^-_l} -\frac{ \gamma^-_l}{\gamma^+_l} \Big( \frac{x^4_h}{x} \Big)^{-\gamma^+_l}\Big] \ind_{\{x< x^4_h\}}
    \end{aligned}
\end{align}

\end{itemize}
\end{proposition}

The above proposition characterises the scenario where, after the health shock, the annuity is more attractive than the financial investment ($\delta_h > \beta_h$) but its purchase involves a fee ($K>0$). In the post-shock regime, the individual will purchase the annuity only when wealth exceeds a critical threshold $x^4_h$, defined in \eqref{xczetac}. This results in the stopping region $\mathcal{S}_h = [x^4_h, \infty)$ (cf. Appendix \ref{APP:VFconstantexplicit}).
The pre-shock strategy depends on the sign of $M^\delta_l$. If $M^\delta_l < 0$, the annuity remains relatively appealing before the shock, and annuitization becomes optimal once wealth rises above a pre-shock threshold $x^4_l$. This threshold is defined as the solution to a nonlinear equation (cf. Proposition \ref{Prop.2}--(i)). The relationship between the pre- and post-shock thresholds, $x^4_l$ and $x^4_h$, generally depends on the model parameters and is not known in closed form. Consequently, the health shock can either shrink or expand the stopping region; that is, the shock may either raise or lower the wealth level required for annuitization to be optimal.
If $M^\delta_l \ge 0$, the annuity is never sufficiently attractive before the shock to outweigh the financial investment, and immediate annuitization is never optimal, i.e., $\cS_l = \varnothing$ (cf. Proposition \ref{Prop.2}--(ii)).

\subsection{Case $K=0$}
The case with no incentives and no fees ($K=0$) leads to a degenerate optimal stopping problem where the optimal strategy is no longer of a threshold type. The following proposition characterises the value function and optimal strategy.
\begin{proposition}
\label{propk0}
Let $K=0$. If $M_l\le 0$ then $\mathcal{C}_l=\varnothing$, $\mathcal{S}_l=[0,\infty)$ and $V(x,\mu_l)=\delta_l x$. If $M_l > 0$ then $\mathcal{C}_l=(0,\infty)$, $\mathcal{S}_l=\{0\}$ and \begin{align}
        V(x,\mu_l)=\Big(\delta_l + \frac{M_l}{r_l+\alpha-\theta} \Big) x.
    \end{align}
\end{proposition}

The above proposition describes the case of no fees or incentives ($K=0$), the annuitization decision is no longer of the threshold type, and the individual either annuitizes immediately or never, depending entirely on the relative attractiveness of the annuity versus the financial investment in each health state.
If, after the health shock, the annuity is at least as attractive as the financial investment ($\delta_h \ge \beta_h$), the individual will choose to annuitize immediately upon the shock's occurrence. The pre-shock strategy, however, depends on the sign of $M^\delta_l$. If $M^\delta_l \le 0$,  the annuity is also attractive in the healthy state, making immediate annuitization optimal. If $M^\delta_l > 0$, the annuity is not appealing before the shock, and the individual will delay annuitization until the health shock occurs.
On the other hand, if the annuity is less attractive than the financial investment after the shock ($\delta_h < \beta_h$), the individual will never annuitize post-shock. The decision in the healthy state is, again, determined by the sign of $M_l^\beta$. If $M_l^\beta \le 0$, annuitization is immediately optimal. If $M_l^\beta > 0$, the annuity is never an attractive option, and the individual will never annuitize.

\section{Numerical analysis of annuitization decisions} \label{numerics}
In this section, we provide a numerical illustration of our theoretical results by comparing the optimal annuitization strategies of two individuals. The first, Individual A, faces a constant mortality force, while the second, Individual B, is exposed to a health shock that increases mortality risk. This comparison highlights how stochastic mortality influences annuitization behaviour. Mortality dynamics are calibrated using data from the \textit{Human Mortality Database}. 
\begin{itemize}
    \item \textbf{Individual A} is a 60-year-old Italian male whose mortality is assumed to follow a constant force of mortality $\mu$ throughout life. To determine this parameter, we rely on the fact that the expected lifetime $\tau_d^A$ for a 60-year-old Italian male in 2010 was 22.41 years. Thus,  \begin{align} \begin{aligned} \E[\tau_d^A]=\int_0^\infty \e^{-\mu t}dt=\frac{1}{\mu}=22.41, \end{aligned}\end{align} which yields $\mu \approx 0.044623$.  

    \item \textbf{Individual B} is also a 60-year-old Italian male and initially shares the same mortality force as Individual A ($\mu_l = \mu$). However, their mortality is subject to a health shock. We set the shock intensity to $\lambda_l = 0.1$, implying the health shock is expected to occur on average after 10 years. Upon occurrence, the mortality force irreversibly jumps to $\mu_h = 1/14.45 \approx 0.069204$, which corresponds to the constant mortality force for a 70-year-old Italian male in 2020. Thus, the increase in the mortality rate due to the health shock is $\Delta = \mu_h - \mu_l \approx 0.024581$. With one million simulations of the path of $\mu_t$, we find that the life expectancy of Individual B is $16.91$ years. 
\end{itemize}

We estimate financial market parameters $\theta$ and $\sigma$ using S{\&}P500 monthly data during the period 1980-2005, obtaining $\theta \approx 0.094864$ and $\sigma \approx 0.154520$.
The objective interest rate $\hat{\rho}$, estimated from 3-month T-bill monthly data over the same period, is $\hat{\rho} \approx 0.059970$, and we set $\rho=\hat{\rho}$. We also fix $\alpha = 0.8\theta \approx 0.075891$ and $\nu=0.25$. 
Then, we assume that the objective mortality force $\hat\mu$ is equal to the constant mortality rate $\mu$ or equivalently to $\mu_l$. Under this assumption, $\delta = 1$ so that the annuity is actuarially fair for Individual A (see Appendix~\ref{APP:VFconstantexplicit}).  
For Individual B, however, fairness is distorted by the health shock. Using \eqref{defdelta0}, we obtain $\delta_l \approx 0.906988$ and $\delta_h \approx 0.809704$. In the high-mortality state, the annuity is overpriced from B’s perspective ($\delta_h < 1$) because they face a higher mortality force $\mu_h$ and the pricing is based on $\hat\mu=\mu_l$. In the low-mortality state, the annuity is also valued below fairness ($\delta_l < 1$), since the individual must account for the possibility of transitioning into the high-mortality state in the future.

Table \ref{tab:parameters} summarises the key parameters used in the numerical implementation of the model.
\begin{table}[H]
\centering
\begin{tabular}{l c l c}
\toprule
\multicolumn{2}{c}{\textbf{Individual Parameters}} & \multicolumn{2}{c}{\textbf{Financial Parameters}} \\
\cmidrule(r){1-2} \cmidrule(l){3-4}
\multicolumn{2}{l}{\textit{Individual A's mortality}} & $\theta$     & $0.094864$ \\
$\mu$        & $0.044623$ & $\sigma$     & $0.154520$ \\
             &            & $\alpha$     & $0.075891$ \\
\addlinespace
\cmidrule(l){3-4}
\multicolumn{2}{l}{\textit{Individual B's mortality}} & \multicolumn{2}{c}{\textbf{Annuity Pricing}} \\
\cmidrule(l){3-4}
$\mu_l$      & $0.044623$ & $\hat{\mu}$  & $0.044623$ \\
$\mu_h$      & $0.069204$ & $\hat{\rho}$ & $0.059970$ \\
$\lambda_l$  & $0.1$      & $K$          &  $-1500$ \\
\addlinespace
\addlinespace
\multicolumn{2}{l}{\textit{Preferences}} & & \\
$\rho$       & $0.059970$ & & \\
$\nu$        & $0.25$     & & \\
\bottomrule
\end{tabular}
\caption{Summary of Model Parameters}
\label{tab:parameters}
\end{table}

\begin{figure}
    \centering
    \includegraphics[width=0.9\linewidth]{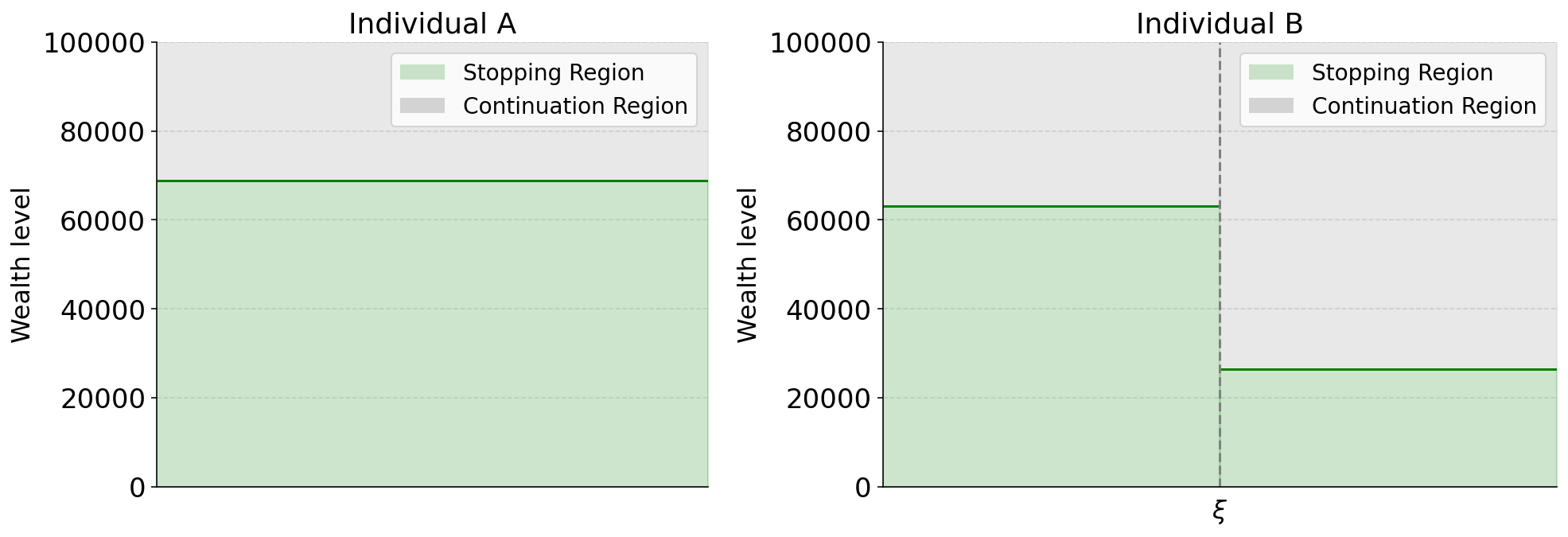}
    \caption{Stopping and continuation regions for Individuals A and B. The green area indicates the stopping region where it is optimal to annuitize, and the grey area represents the continuation region where individuals postpone annuitization. The left panel shows Individual A's optimal annuitization threshold of $x^2 = 68893.49$. The right panel shows Individual B's optimal strategy. Before the shock, annuitization is optimal below $x^2_l = 63132.55$, while after the shock (at time $\xi$), the threshold decreases to $x^2_h = 26431.37$.}
    \label{fig:thresholds}
\end{figure}

We next examine optimal annuitization strategies. For Individual A, following \cite{buttarazzi2024market} (cf. also Appendix~\ref{APP:VFconstantexplicit}), since $\delta \le \beta$ and $K < 0$, the continuation region is $(x^2, \infty)$ and the stopping region is $[x^2,\infty]$ with $x^2 = 68893.49$. 
In economic terms, Individual A chooses to annuitize when wealth is relatively low. Purchasing the annuity transfers longevity risk to the insurer and guarantees a lifelong income stream, which is particularly valuable when liquid wealth may be insufficient to sustain desired consumption. Conversely, when wealth is sufficiently high (exceeding the threshold $x^2$), the individual prefers to keep wealth invested in the financial market. This behaviour is also influenced by bequest motives: individuals who wish to leave wealth to heirs may delay annuitization to preserve potential inheritance, especially given the irreversibility of the decision.
The strategy for Individual B differs depending on whether the health shock has occurred. In the post-shock state, we find $\delta_h \ge \beta_h$ implying that the continuation region is $(x^2_h, \infty)$ and the stopping region is $[0,x^2_h]$ with $x^2_h = 26431.37$.  
In the pre-shock state, since $M^\delta_l > 0$, Proposition~\ref{Klowerbeta} implies that the continuation and stopping regions are $\mathcal{C}_l=(x^2_l,\infty)$ and $\mathcal{S}_l=[0,x^2_l]$ where $x^2_l = 63132.55$.
The presence of two stopping regions for Individual B reflects the stochastic nature of their mortality. Before the health shock, the individual faces the same mortality rate as Individual A, but the pre-shock stopping threshold is slightly lower ($x^2_l < x^2$) because they anticipate the possibility of future health deterioration. After the shock, the threshold decreases further: individuals who delay annuitization until after the shock face both a shorter expected lifespan and a higher mortality rate, which reduces the attractiveness of converting wealth into an annuity.
Figure~\ref{fig:thresholds} summarizes the above discussion by illustrating the differences in stopping regions and annuitization thresholds for Individuals A and B across mortality states.

We simulate the wealth dynamics of both individuals to examine how the annuitization thresholds translate into actual purchase behaviour. We generate 100,000 potential wealth trajectories over a 20-year horizon, from age 60 to 80, assuming an initial wealth of 100,000 USD. The simulation uses a time step of $dt= 3.97 \cdot 10^{-3}$, corresponding to daily updates of the wealth process (approximately 252 trading days per year).
For each path, the shock time $\xi$ is simulated from an exponential distribution with intensity $\lambda_l$.
For Individual A, the simulation shows that roughly 51.9\% of paths cross the annuitization threshold $x^2$, reflecting a relatively high probability of purchasing the annuity. Conditional on purchase, the average crossing time is 6.84 years.
For Individual B, the stochastic health shock significantly affects annuitization behaviour. Only 21.7\% of paths result in annuitization before the shock, whereas only 1.3\% annuitize after the shock. The average waiting time until purchase is slightly shorter at 6.48 years.
This drastic reduction in post-shock annuitization is a direct consequence of the threshold dropping from $x^2_l$ to $x^2_h$. Had the pre-shock threshold $x^2_l$ remained in place, a much larger fraction of wealth paths would have crossed it, producing annuitization probabilities closer to those of Individual A. However, once the health shock occurs, the new threshold $x^2_h$ is so low that the majority of wealth trajectories are very unlikely to reach it, effectively shutting down the option to annuitize after the health deterioration.

\subsection{Sensitivity to the size of health shock}
We now conduct a sensitivity analysis to examine the influence of the size of the health shock $\Delta$ on Individual B's optimal strategy. When the shock occurs, the mortality force increases irreversibly from $\mu_l$ to $\mu_h(\Delta) = \mu_l + \Delta$, with $\Delta$ ranging from $0$ up to $\Delta_{\text{max}} = 0.229350$. The maximum shock sets the post-shock mortality $\mu_h(\Delta_{\text{max}})$ to that of a 90-year-old Italian male in 2020 (with a remaining life expectancy of 3.65 years).

\begin{figure}[h]
    \centering
    \begin{subfigure}[t]{0.45\textwidth}
        \centering
        \includegraphics[width=0.9\textwidth]{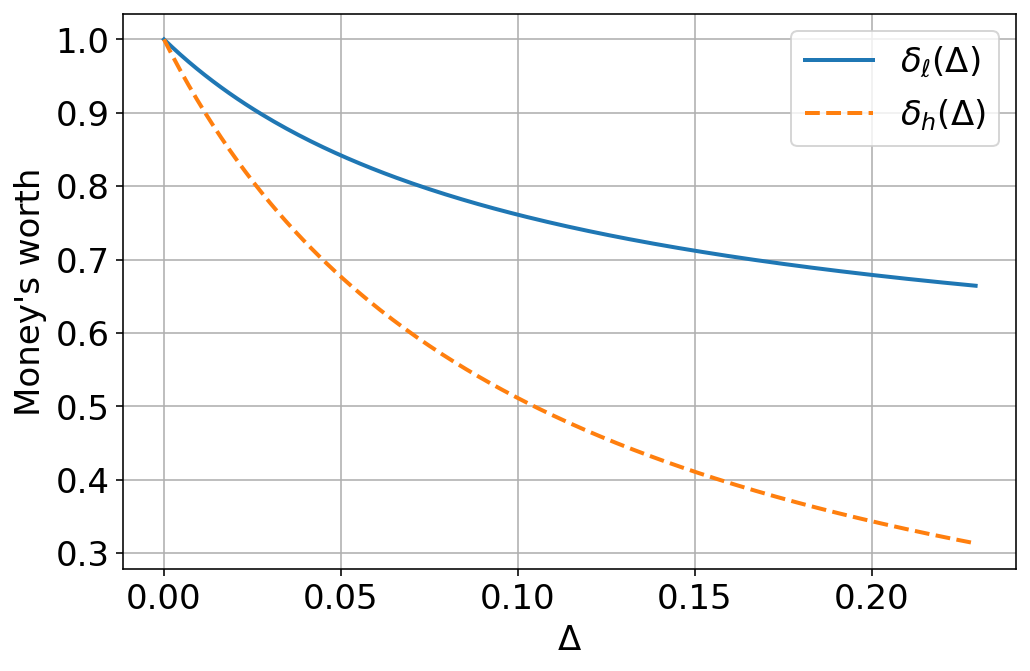}
        \caption{}
        \label{fig:delta_l_h}
    \end{subfigure}
    \hfill
    \begin{subfigure}[t]{0.45\textwidth}
        \centering
        \includegraphics[width=0.9\textwidth]{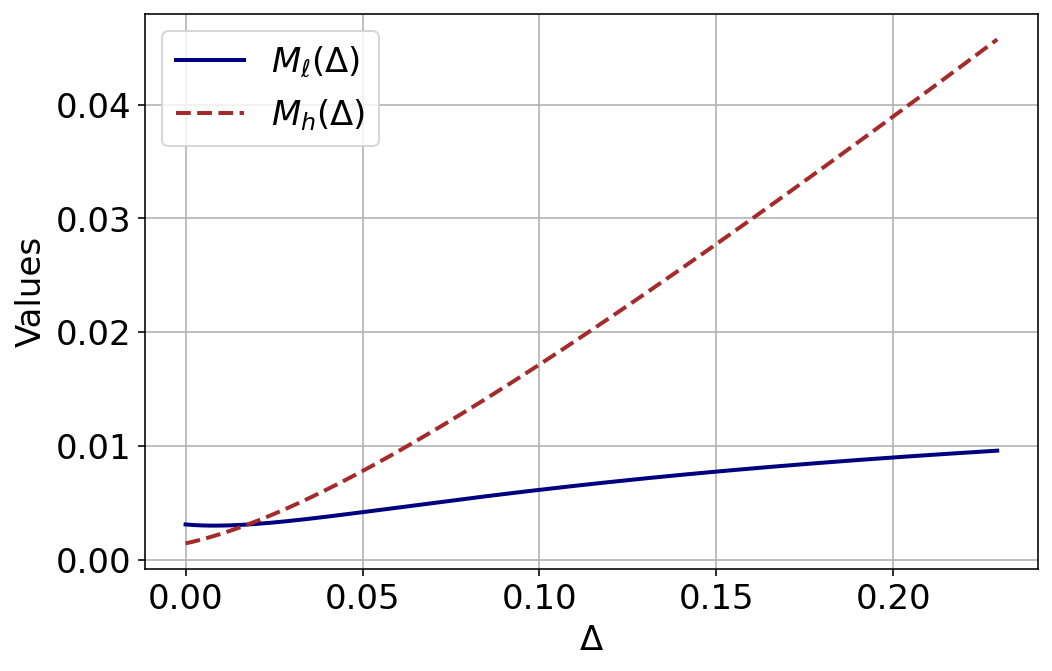} 
        \caption{}
        \label{fig:Mplots}
    \end{subfigure}

    \vspace{0.5cm} 

    \begin{subfigure}[b]{0.45\textwidth} 
        \centering
        \includegraphics[width=0.9\textwidth]{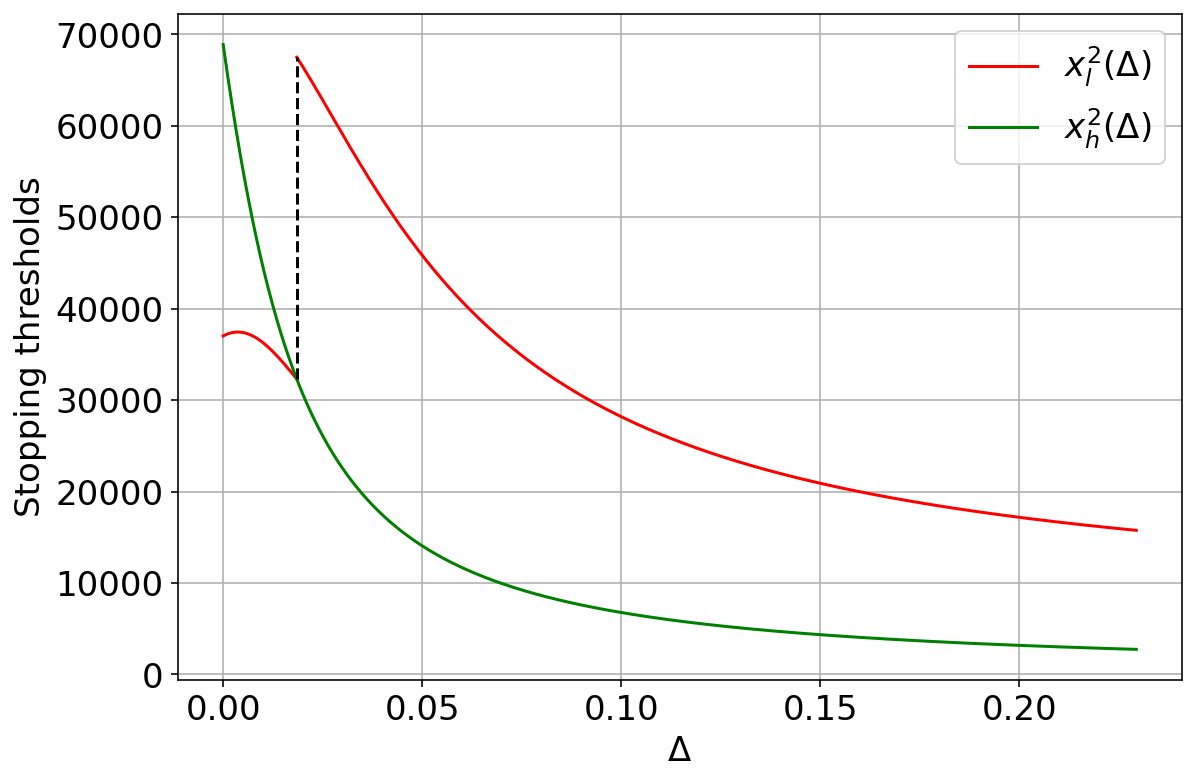}
        \caption{}
        \label{fig:x2lh}
    \end{subfigure}
    \caption{(\subref{fig:delta_l_h}) The annuity's money's worth coefficients $\delta_i(\Delta)$ in the pre-shock (blue solid line) and post-shock (yellow dashed line) states. (\subref{fig:Mplots}) The index $M_i(\Delta)$ in the pre-shock (blue navy line) and post-shock (brown dashed line) states. (\subref{fig:x2lh}) Optimal annuitization thresholds $x^2_l(\Delta)$ (red line) and $x^2_h(\Delta)$ (green line).}
\end{figure}

The perceived fairness of the annuity, from Individual B's perspective, is a key driver of behaviour and is directly affected by $\Delta$ (see Equation \eqref{defdelta0}). In particular, if $\Delta = 0$, then $\delta_l(\Delta) = \delta_h(\Delta) = 1$, and the annuity remains fair. By contrast, if $\Delta > 0$, both $\delta_l(\Delta)$ and $\delta_h(\Delta)$ decrease monotonically in $\Delta$, implying that the annuity becomes overpriced and hence less attractive from Individual B’s perspective. Figure \ref{fig:delta_l_h} shows $\delta_l(\Delta)$ and $\delta_h(\Delta)$ as functions of $\Delta$, illustrating that $\delta_h(\Delta)$ declines more rapidly than $\delta_l(\Delta)$.
This difference in sensitivity can be explained analytically. From the expression \eqref{defdelta0}, we find the relationship 
\[
\bigg|\frac{\partial \delta_h(\Delta)}{\partial \Delta}\bigg| = C \,\bigg|\frac{\partial \delta_l(\Delta)}{\partial \Delta}\bigg|,
\]
with $C\coloneqq(\rho+\lambda_l+\mu_l)/ \lambda_l$.
The magnitude of $C$ reflects both the intensity of the health shock $\lambda_l$ and the effective discount rate $\rho+\lambda_l+\mu_l$. Since $C>1$, increases in $\Delta$ affect $\delta_l(\Delta)$ less strongly than $\delta_h(\Delta)$. Intuitively, the pre-shock value $\delta_l(\Delta)$ incorporates the probability of transitioning to the post-shock regime, which partially absorbs the impact of higher post-shock mortality. By contrast, $\delta_h(\Delta)$ is based solely on the elevated mortality force $\mu_h$, explaining its sharper decline.

The coefficients $M_i$ defined in \eqref{M_i} provide a useful summary measure of the attractiveness of annuitization in each mortality regime. Intuitively, a smaller $M_i$ indicates a stronger incentive to annuitize rather than remain invested in the financial market in the heath state $i\in\{h,l\}$. Figure~\ref{fig:Mplots} reveals a critical shift: for small shocks ($\Delta < 0.01755$), we find $M_h(\Delta) < M_l(\Delta)$, indicating that annuitization is more attractive after the shock has occurred than before. For larger shocks ($\Delta > 0.01755$), this inequality reverses, making annuitization more attractive before the shock.
Figure~\ref{fig:x2lh} illustrates the resulting stopping thresholds, $x^2_l(\Delta)$ (pre-shock) and $x^2_h(\Delta)$ (post-shock). The post-shock threshold $x^2_h(\Delta)$ is monotonically decreasing in $\Delta$; a larger mortality shock shortens life expectancy and makes the annuity less valuable, so a lower wealth level is required to trigger the purchase.
The behaviour of the pre-shock threshold $x^2_l(\Delta)$ is more complex and non-monotonic, leading to two distinct regimes based on the relative position with $x^2_h(\Delta)$.
\begin{itemize}
\item For $\Delta \le 0.01849$, $x^2_l(\Delta) \le x^2_h(\Delta)$. The potential cost of buying the annuity later at a slightly worse price is outweighed by the opportunity for higher financial returns, making the individual more willing to forgo annuitization. After the shock, mortality is only slightly elevated. The annuity, while somewhat less valuable, remains relatively attractive. Since the shock typically occurs after about ten years, individuals in this post-shock state face a shorter horizon and are less willing to take on financial risk.
\item For $\Delta > 0.01849$, $x^2_l(\Delta) > x^2_h(\Delta)$. In this region, $M_l(\Delta) > M_h (\Delta)$, indicating that annuitization is relatively more attractive before the shock. Here, individuals who did not annuitize before the shock now have less time to benefit from annuity payments, reducing their willingness to purchase. After the shock, the annuity becomes substantially less attractive due to the drastically shortened life expectancy, further discouraging late annuitization. 
\end{itemize}

\subsection{Sensitivity to the intensity of the health shock}

We now analyse how the intensity of the health shock, $\lambda_l$, influences the optimal annuitization strategy for Individual B. This parameter governs the expected arrival time of the shock, with a higher $\lambda_l$ indicating a greater probability of health deterioration in the near term. 
We notice that the post-shock environment is unaffected by the shock’s intensity: the parameters $\delta_h$, $\beta_h$, and the post-shock annuitization threshold $x^2_h$ do not depend on $\lambda_l$.

\begin{figure}[h]
    \centering
    \begin{subfigure}[t]{0.45\textwidth}
        \centering
        \includegraphics[width=0.9\textwidth]{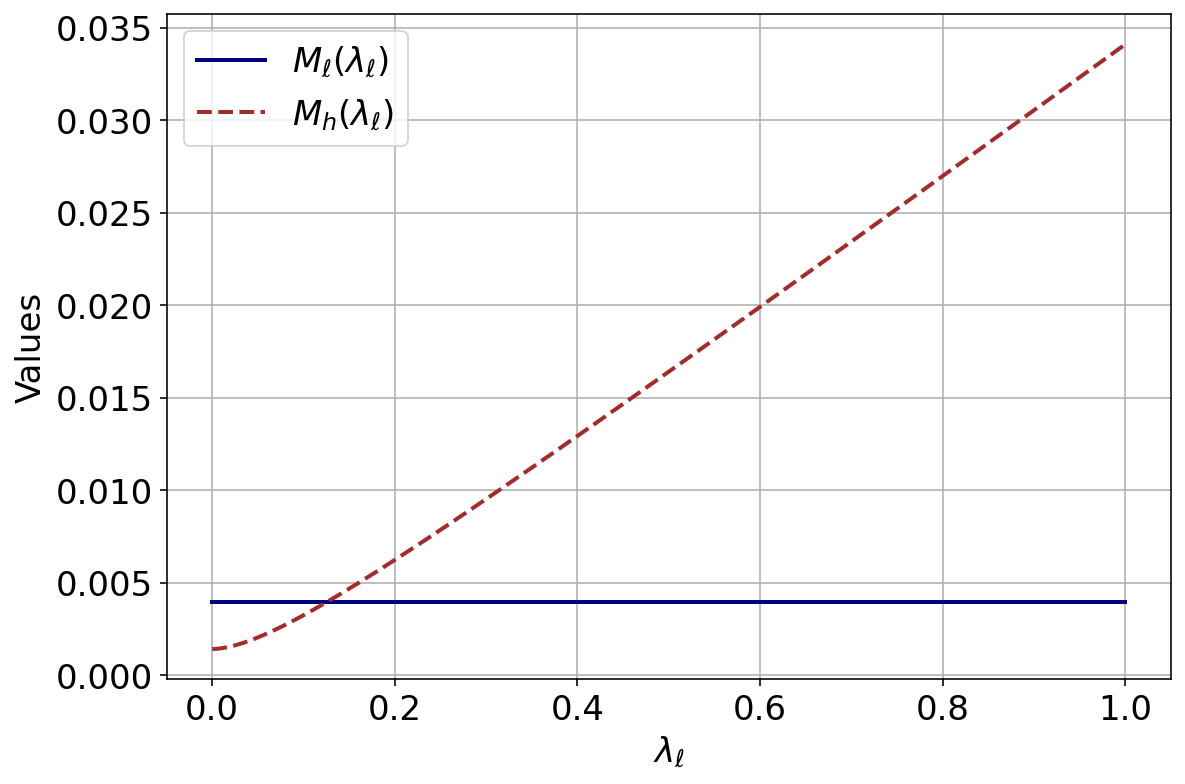}
        \caption{}
        \label{fig:Mplotslambda}
    \end{subfigure}
    \hfill
    \begin{subfigure}[t]{0.45\textwidth}
        \centering
        \includegraphics[width=0.9\textwidth]{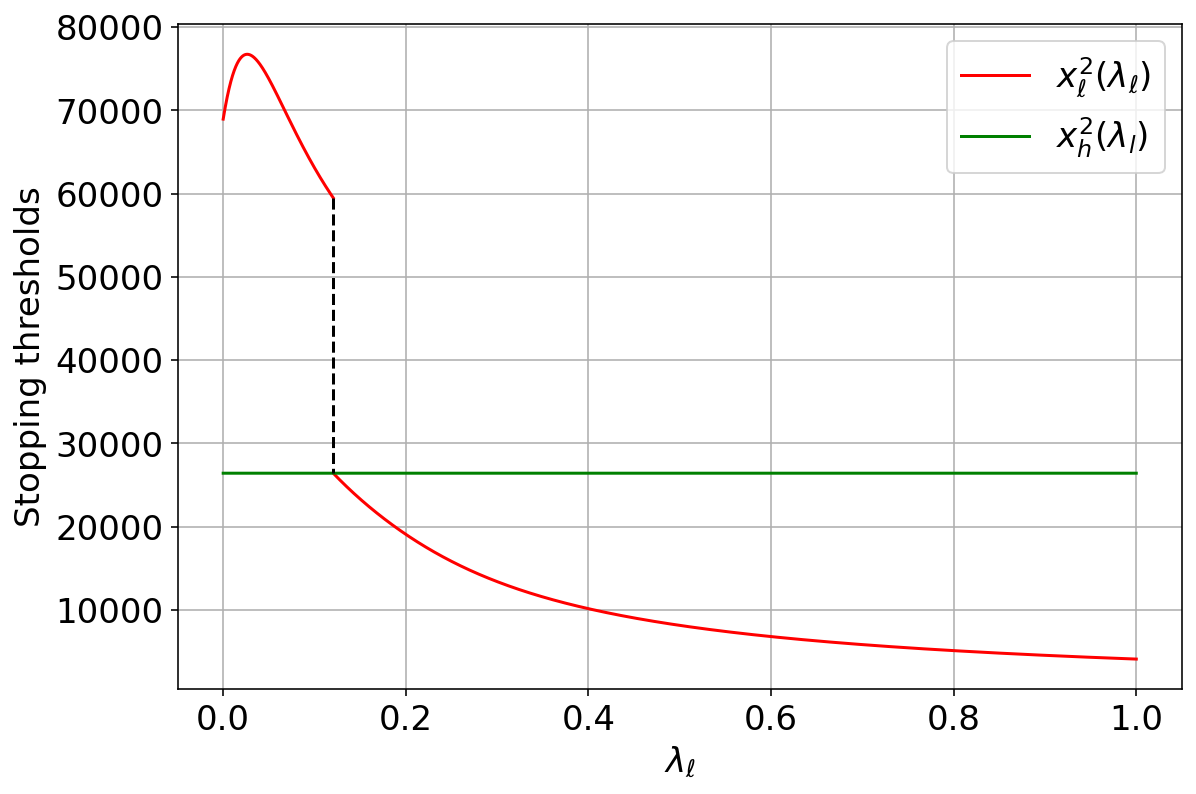} 
        \caption{}
        \label{fig:x2lhlambda}
    \end{subfigure}

    \caption{(\subref{fig:Mplotslambda}) The index $M_i(\lambda_l)$ in the pre-shock (navy solid line) and post-shock (brown dashed line) states. (\subref{fig:x2lhlambda}) Optimal annuitization thresholds $x^2_l(\lambda_l)$ (red line) and $x^2_h(\lambda_l)$ (green line).}
\end{figure}

The indices $M_i(\lambda_l)$ again provide insight into the relative attractiveness of annuitization across regimes. Figure~\ref{fig:Mplotslambda} shows that when $\lambda_l$ is sufficiently low ($\lambda_l < 0.125210$), we have $M_h(\lambda_l)<M_l(\lambda_l)$, meaning that annuitization is relatively more attractive in the post-shock regime. For higher intensities, the inequality reverses, and pre-shock annuitization becomes more attractive.
Figure~\ref{fig:x2lhlambda} illustrates the corresponding stopping thresholds. The post-shock threshold $x^2_h$ is constant in $\lambda_l$, while the pre-shock threshold $x^2_l(\lambda_l)$ exhibits a more complex, non-monotonic pattern. When $\lambda_l$ is small (below $0.12063$), the shock is expected to occur late, implying a longer expected lifetime in the pre-shock state. This makes annuitization appealing before the shock, since the individual can expect to receive annuity payments over a longer horizon, and we observe $x^2_l(\lambda_l)>x^2_h$. By contrast, when $\lambda_l$ is large, the shock is likely to arrive soon, shortening the duration of the pre-shock state. In this case, annuitization before the shock becomes less attractive, and the ordering reverses with $x^2_l(\lambda_l)\le x^2_h$. In line with this behaviour, for high intensities $M_h(\lambda_l)>M_l(\lambda_l)$, confirming that annuitization is relatively more attractive in the post-shock regime.

\section{Conlusions}
This study examines the optimal annuitization decision for an individual exposed to the risk of a sudden and permanent health deterioration. We derive explicit analytical characterizations of the value function and the optimal stopping boundaries, which are organized into distinct cases based on model parameters (Propositions \ref{propKlowerDelta}--\ref{propk0}). These results illustrate the sophisticated trade-off an individual must navigate when balancing financial risks against longevity risks.

Our numerical analysis reveals that for an individual facing health shock risk (Individual B), the annuitization decision involves a nuanced assessment of both financial factors and the evolution of health status. This is in contrast to the simpler, wealth-based rule for an individual with constant mortality (Individual A). 
The sensitivity analyses demonstrate that optimal behaviour can be categorized into distinct regimes: when a shock is either small or expected far in the future, annuitization is relatively more attractive in the post-shock regime. Conversely, when a shock is large and/or imminent, the individual has a stronger incentive to annuitize before their health deteriorates.
In summary, our model provides a formal framework for determining the optimal annuitization exercise rule, offering valuable insights for both retirement planning and annuity product design.

Several directions remain for future research. First, the model could be generalised to include multiple, non-absorbing health states to capture a wider range of health trajectories, such as the possibility of recovery. Second, relaxing the all-or-nothing annuitization assumption to allow for partial purchases would significantly enhance the model's practical relevance. Finally, introducing dependencies between health status and other model parameters—for example, allowing the health shock to affect risk aversion or bequest motives—would create a more realistic framework for assessing retirement decisions under interconnected risks.

\subsection*{Conflicts of interest.} The author has no relevant financial or non-financial interests to disclose.

\appendix
\section{Value function for the constant mortality force problem}
\label{APP:VFconstantexplicit}

In this appendix, we recall from \cite{buttarazzi2024market} the explicit expressions for the value function and the associated continuation and stopping sets for the problem with a constant mortality force $\mu_t = \mu$ for all $t \ge 0$. The results are presented for a generic mortality force $\mu$. This allows us to recover the specific case for the high-mortality state ($h$) by simply substituting $\mu = \mu_h$. In this case, all resulting quantities ($\beta$, $\delta$, $\gamma^\pm$, $x^2$, $x^4$, $\zeta^2$, $\zeta^4$, $\mathcal{S}$, $\mathcal{C}$) then become the $h$-indexed constants and sets used in the main text.
We remark that \cite{buttarazzi2024market} studies the optimal stopping problem on the state space $(0,\infty)$, whereas \cite{buttarazzi2025optimal} extends the analysis to $[0,\infty)$. Hence, by applying \cite[Thm.~2.7]{buttarazzi2025optimal}, we present the continuation and stopping regions in the enlarged state space.

Let $\theta-\alpha-\rho-\mu<0$ and define
\begin{align}
    \beta \coloneqq \frac{\alpha + \nu \mu}{\theta - \alpha - \rho - \mu}, 
    \qquad 
    \delta \coloneqq \frac{\hat \rho + \hat \mu}{\rho + \mu},
\end{align}
and
\begin{equation} 
\gamma^{\pm}\coloneqq \frac{1}{2} - \frac{\theta - \alpha}{\sigma^2} \pm \sqrt{\left( \frac{1}{2} - \frac{\theta - \alpha}{\sigma^2} \right)^2 + \frac{2 (\rho + \mu)}{\sigma^2}}.
\end{equation}
Define 
\begin{align} 
\mathcal{S}\coloneqq\{x\in [0,\infty): V(x,\mu)= \delta (x-K)\}, \qquad \mathcal{C} \coloneqq[0,\infty)\setminus \mathcal{S}.
\end{align}
Then, we have the following cases.

\begin{itemize}
    \item[\textnormal{(1)}]  Let $K > 0$.
    \begin{itemize}
    \item[i)] If $\delta\geq \beta$, then $\mathcal{C}=\varnothing$, $\mathcal{S}=[0,\infty)$ and \begin{equation} \label{V0knegdelta1bigger} V(x,\mu)= \delta(x-K).\end{equation}
    
    \item[ii)] If $\delta<\beta$, then $\mathcal{C}=(x^2,\infty)$, $\mathcal{S}=[0,x^2]$ and
    \begin{equation}  V(x,\mu)=\begin{cases} \label{V0knegdelta1lower} \beta x + \zeta^2 x^{\gamma^-}, \hspace{5mm} &\text{ if } x>x^2,\\
    \delta (x-K), \hspace{5mm} &\text{ if } x\leq x^2,\\
    \end{cases}\end{equation}
    with 
    \begin{align}
    \begin{aligned}
    x^2\coloneqq \frac{\delta K \gamma^-}{(\gamma^--1)(\delta-\beta)}, \qquad
    \zeta^2\coloneqq \Big(\frac{\delta K}{\gamma^--1}\Big)^{1-\gamma^-}\Big(\frac{\delta - \beta}{\gamma^-}\Big)^{\gamma^-}.
    \end{aligned}
    \end{align}
    \end{itemize}
    
\item[\textnormal{(2)}]  Let $K < 0$.
    \begin{itemize}
    \item[i)] If $\delta\leq \beta$, then $\mathcal{C}=[0,\infty)$, $\mathcal{S}=\varnothing$ and \begin{equation} \label{V0kposdelta1lower} V(x,\mu)=\beta x. \end{equation} 
    
    \item[ii)] If $\delta>\beta$, then $\mathcal{C}=[0,x^4)$, $\mathcal{S}=[x^4,\infty)$ and
    \begin{equation}\label{V0kposdelta1bigger}  V(x,\mu)=\begin{cases} \beta x + \zeta^4 x^{\gamma^+}, \hspace{5mm} &\text{ if } x<x^4,\\
    \delta (x-K), \hspace{5mm} &\text{ if } x\geq x^4,\\
    \end{cases}\end{equation}
    with
    \begin{align} 
    \begin{aligned}
    x^4\coloneqq \frac{\delta K \gamma^+}{(\gamma^+-1)(\delta-\beta)} ,\qquad    \zeta^4\coloneqq \Big(\frac{\delta K}{\gamma^+-1}\Big)^{1-\gamma^+}\Big(\frac{\delta - \beta}{\gamma^+}\Big)^{\gamma^+}.
    \end{aligned}
    \end{align}
    \end{itemize}   
    \item[\textnormal{(3)}]  Let $K = 0$. If $\delta < \beta$, then $\mathcal{C}=(0,\infty)$, $\mathcal{S}=\{0\}$ and $V(x,\mu)=\beta x$. If $\delta\geq \beta$, then $\mathcal{C}=\varnothing$, $\mathcal{S}=[0,\infty)$ and $V(x,\mu)= \delta x$.
\end{itemize}

\bibliographystyle{plain}
\bibliography{biblio}

\end{document}


\maketitle

This supplementary material provides the detailed proofs for the results presented in the main text. It is structured in two sections. Section \ref{moneyworthsection} contains the proof of Proposition 2.2, which derives the explicit expression for the money's worth ratio. Section \ref{sec:explicit_solution} presents the analytical foundation and proofs for the main results of the paper (Propositions 3.2–3.6 of the main text).

\section{Money's worth} \label{moneyworthsection}
\begin{proof}[\textbf{Proof of Proposition 2.2 in the main text}]
Using the independence between $\cF^B_\infty$ and $\cH_\infty$ in Assumption 2.1 of the main text, and applying Fubini's theorem, we get \begin{align} \begin{aligned} \label{afub}
    \frac{a_{t}}{\hat a_t}= (\hat\rho+\hat\mu) 
 \int_{0}^{\infty} \e^{-\rho u} \E \Big[ \e^{ -\int_{t}^{t+u}\mu_s ds}\ \Big| \ \cH_t \Big] du.
\end{aligned}\end{align}
From  \cite[Proposition 1]{duffie2000transform}, one has
\begin{equation} 
\label{Eq:Duffie}
\E \Big[ \e^{ -\int_{t}^{T} \mu_s ds}\Big|\cH_t \Big]= e^{\alpha(t)+\beta(t)H_t}=\ind_{\{t<\xi\}} e^{\alpha(t)}+ \ind_{\{t\geq \xi\}} e^{\alpha(t)+\beta(t)},
\end{equation}
where the coefficients $\alpha(\cdot)$ and $\beta(\cdot)$ satisfy 
\begin{equation}
\begin{cases} \label{ODEsalphabeta}
    \Dot{\alpha}(t)=\mu_l-\lambda_l[e^{\beta (t)}-1],\\
    \Dot{\beta}(t)=\mu_h-\mu_l +\lambda_l[e^{\beta (t)}-1],
\end{cases}
\end{equation}
with the boundary conditions $\alpha(T)=0$ and $\beta(T)=0$. For $\Delta\mu - \lambda \neq 0$, the explicit solutions of the ODEs \eqref{ODEsalphabeta} are
\begin{align}
\begin{aligned}
    \alpha(t)&=  -\mu_h  (T-t)  -\beta(t),\\
    \beta(t)&= - \ln \Big(\frac{\Delta\mu}{\Delta\mu -\lambda} \e^{(\Delta\mu -\lambda)(T-t)}- \frac{\lambda}{\Delta\mu -\lambda} \Big).
\end{aligned}
\end{align}
Whereas for $\Delta\mu - \lambda = 0$, the explicit solutions are
\begin{align}
\begin{aligned}
\alpha(t) &=- \mu_h (T - t) - \beta(t),\\
\beta(t) &= -\ln\big(1 + \lambda (T - t)\big).
\end{aligned}
\end{align}
Combining the explicit expressions for $\alpha(\cdot)$ and $\beta(\cdot)$ with \eqref{Eq:Duffie} and setting $T=t+u$, we find
\begin{itemize}
    \item[i)] For $\Delta\mu - \lambda \neq 0$, 
        \begin{align}
        \begin{aligned} 
            \E_\pi \bigg[ \e^{ -\int_{t}^{t+u} \mu_s \ud s} \Big| \cH_t \bigg]&= \ind_{\{t<\xi\}} \Big( \frac{\Delta\mu }{\Delta\mu -\lambda}\e^{-(\mu_l+\lambda)u} -  \frac{\lambda}{\Delta\mu-\lambda}\e^{-\mu_h u} \Big)  + \ind_{\{t\geq \xi\}} \e^{-\mu_h u} 
        \end{aligned}
        \end{align}  
    \item[ii)] For $\Delta\mu - \lambda = 0$, 
        \begin{align}
        \begin{aligned} 
        \E_\pi \bigg[ \e^{ -\int_{t}^{t+u} \mu_s \ud s} \Big| \cH_t \bigg]&= \ind_{\{t<\xi\}} \big(1+\lambda u\big) \e^{-\mu_h u} + \ind_{\{t\geq \xi\}} \e^{-\mu_h u} 
        \end{aligned}
        \end{align} 
\end{itemize}

Combining these expressions with \eqref{afub}, and observing that 
\begin{align}
    &\ind_{\{\Delta\mu\neq \lambda\}} \int_0^\infty \Big( \frac{\Delta\mu}{\Delta\mu-\lambda} \e^{-(\rho+\mu_l+\lambda)t} -  \frac{\lambda}{\Delta\mu-\lambda} \e^{-(\rho+\mu_h)t} \Big) \ud t + \ind_{\{\Delta\mu= \lambda\}} \int_0^\infty \big(1+\lambda t \big) \e^{-(\rho+\mu_h)t} \ud t\\
    &= \ind_{\{\Delta\mu\neq \lambda\}} \Big(\frac{\Delta\mu}{(\Delta\mu-\lambda)(\rho+\mu_l+\lambda)} - \frac{\lambda}{(\Delta\mu-\lambda)(\rho+\mu_h)} \Big) +  \ind_{\{\Delta\mu= \lambda\}} \frac{1}{\rho+\mu_h} \Big(1 + \frac{\lambda}{\rho+\mu_h}\Big)\\
    &= \frac{\rho+\mu_l+\Delta\mu +\lambda}{(\rho+\mu_l+\lambda)(\rho+\mu_l+\Delta\mu)},
\end{align}
we conclude the proof.
\end{proof}

\section{Explicit Solution} \label{sec:explicit_solution}
In this section, we derive the explicit solution for the value function $V(x,\mu_l)$ and characterise the associated continuation and stopping sets, thereby proving the main results (Propositions 3.2--3.6) of the paper. In Subsection \ref{Analytical}, we recall essential theoretical results from \cite{buttarazzi2025optimal} that form the foundation for our analysis. Then, in Subsection \ref{vfmuh}, we recall the explicit solution for the post-shock problem with constant mortality force $\mu_t = \mu_h$. Finally, Subsection \ref{proofresultsmain} contains the proofs of the main propositions.

To make this supplementary material more self-contained, we collect here some of the key constants from the main text. For $i\in\{l,h\}$ the parameters $\delta_i$ and $\beta_i$ are given by
\begin{align} \label{recallingdeltabeta}
\delta_i = \frac{(\rho + \lambda_i + \mu_h)(\hat{\rho}+\hat{\mu})}{(\rho + \lambda_i + \mu_i)(\rho+\mu_h)}, \qquad \qquad
\beta_i = \frac{\alpha+\nu \mu_i}{r_i+\alpha-\theta},
\end{align}
and $r_i=\rho+\mu_i+\lambda_i$ with $\lambda_h=0$. Moreover, the parameter $M_i$ is given by \begin{align}
    M_i=(\delta_i-\beta_i)(\theta-\alpha-r_i ) +\lambda_i \max\{\delta_h,\beta_h\}
\end{align} and we recall for the low-mortality state
\begin{align}
\label{recallingMldeltabeta}
M^\delta_l = \ind_{\{\delta_h\ge \beta_h\}} M_l, \qquad  M^\beta_l   = \ind_{\{\delta_h < \beta_h\}} M_l.
\end{align}
Finally, the constants $\gamma^{\pm}_i$ are defined in the main text as 
\begin{equation} \label{recallgamma}
\gamma^{\pm}_i = \frac{1}{2} - \frac{\theta - \alpha}{\sigma^2} \pm \sqrt{\left( \frac{1}{2} - \frac{\theta - \alpha}{\sigma^2} \right)^2 + \frac{2 r_i}{\sigma^2}}.
\end{equation}
Notice that $\gamma^+_i>1$ and $\gamma^-_i<0$.

\subsection{Analytical foundation} \label{Analytical}
In this subsection, we recall theoretical results from \cite{buttarazzi2025optimal}. We begin by introducing the so-called \emph{Lagrange formulation} of the problem. Following \cite[Eqs. (2.10)--(2.12)]{buttarazzi2025optimal}, we define
 \begin{align} \label{ValuefunctionWonejump}
W(x,\mu_l)\coloneqq V(x,\mu_l) - \delta_l (x-K).
\end{align}
An application of Dynkin's formula to $\e^{-r_l \tau} \delta_l (X_\tau - K)$ then yields the representation
\begin{align}
\begin{aligned}
W(x,\mu_l)= \sup_{\tau \in \cT(\bF^B)} \E_{x,\mu_l} \bigg[ \int_0^\tau e^{-r_l  t}  M(X_t,\mu_l) \ud t  \bigg],
\end{aligned}\end{align} 
with 
\begin{align} \begin{aligned}
\label{Monejump} 
M(x,\mu_l)&\coloneqq r_l \delta_l K + \big(\delta_l-\beta_l\big)(\theta-\alpha-r_l)x+\lambda_l V(x,\mu_h).
\end{aligned}\end{align} 
Notice that the value function $W(x,\mu_l)$ (and $V(x,\mu_l)$) depends explicitly on $\mu_l$ in a parametric way. Thus, we can also replace $\E_{x,\mu_l}$ with the easier $\E_x$. Thanks to the explicit form of the process $X$, we will often use the equivalence between the law of $X$ under $\P_x$ and the law of $X^x$ under $\P$.

Next, we present sufficient conditions for characterising the structure of the continuation and stopping regions. To maintain a concise presentation, we only include the essential criteria required for our analysis from \cite[Theorems 2.6 and 2.7]{buttarazzi2025optimal}. Furthermore, we adapt these theorems to our specific context and notation for clarity and self-containedness.

Recall the infinitesimal generator $\operL$ of the process $(X_t)_{t\geq 0}$
\begin{align}\label{L}
(\cL u)(x)=\tfrac{1}{2} \ \sigma^2 x^2 u_{xx}(x)+(\theta-\alpha)x u_{x}(x), \quad \text{for any} \ u\in C^2(\R),
\end{align}
where $u_x=\partial u/\partial x$ and $u_{xx}=\partial^2 u/\partial x^2$. The following theorem describes the regularity of the value function and introduces the free-boundary problem that we will solve to obtain closed-form expressions for the value function.

\begin{theorem}[Theorem 2.6 from \cite{buttarazzi2025optimal}]
\label{regularityvf}
The mapping $x\mapsto V(x,\mu_l)$ is Lipschitz and {\em continuously differentiable} on $(0,\infty)$ (in particular, $V(\cdot,\mu_l)$ extends continuously to $[0,\infty)$). Moreover, $V$ is a classical solution of the free-boundary problem
\begin{align}\label{eq:ODE}
\begin{aligned}
\operL V(x,\mu_l) -r_l V(x,\mu_l)=-(\alpha+\nu\mu_l)x-\lambda_l V(x,\mu_h), \quad x\in\cC_l,
\end{aligned}
\end{align}
and
$V(\cdot,\mu_l)\in C^2(\overline\cC_l \cap(0,\infty))$, where $\overline\cC_l$ denote the closure of the continuation set $\cC_l$ relative to $[0,\infty)$.
\end{theorem}

It is also useful to express the above free-boundary problem in terms of the \textit{Lagrange reformulation} of the value function $W(x,\mu_l)$, which becomes
\begin{align}\label{BPVEq}
\begin{alignedat}{2}
    (\operL-r_l) W(x,\mu_l)&= M(x,\mu_l), \quad && x\in\mathcal{C}_l,\\
    W(x,\mu_l)&= 0, \quad && x\in\mathcal{S}_l,\\
    W_x(x,\mu_l)&= 0, \quad && x\in\partial \mathcal{C}_l.
\end{alignedat}
\end{align}

Next, the following theorem provides criteria to identify the optimal annuitization rules. 

\begin{theorem}[Theorem 2.7 from \cite{buttarazzi2025optimal}]
\label{proposuffcondm}
The structure of $\mathcal{C}_{l}$ and $\mathcal{S}_{l}$ is determined as follows:
\begin{itemize}
    \item[\textnormal{(1)}]  Let $K > 0$.
    \begin{itemize}
    \item[i)] If $M(x,\mu_l)>0$ for all $x\in [0,\infty)$, then $\mathcal{C}_l=[0,\infty)$ and $\mathcal{S}_l=\varnothing$.
    \item[ii)] If $M(x,\mu_l)\to-L$ as $x\to\infty$ for some $L\in(0,\infty]$, then there is some $b>0$ such that $\mathcal{C}_l=[0,b)$ and $\mathcal{S}_l=[b,\infty)$. 
    \end{itemize} 

    \item[\textnormal{(2)}] Let $K < 0$.
    \begin{itemize}
    \item[i)] If $M(x,\mu_l)\leq0$ for all $x\in [0,\infty)$, then $\mathcal{C}_l=\varnothing$ and $\mathcal{S}_l=[0,\infty)$.
    \item[ii)] If there exists $\hat{x} \in [0,\infty)$ such that $M(\hat{x},\mu_l)>0$, then there is some $b>0$ such that $\mathcal{C}_l=(b,\infty)$ and $\mathcal{S}_l=[0,b)$. 
    \end{itemize}

    \item[\textnormal{(3)}] Let $K = 0$. Then $M(x,\mu_l)=M_l \cdot x$. Moreover, $\{0\}\in\cS_{l}$ whereas $M_l \le 0\implies\cC_{l}=\varnothing$ and $M_l>0\implies \cC_{l}=(0,\infty)$. 
\end{itemize}
\end{theorem}

Finally, we recall \cite[Proposition 5.1]{buttarazzi2025optimal}.
\begin{proposition} 
\label{Vnfiniteness}
Under assumption 2.3 of the main text, there exists $L>0$, independent of $(x,\mu_l)$, such that $0\le V(x,\mu_l)\leq L(1+x)$ for $x\in(0,\infty)$.
\end{proposition}

The shape of the free-boundary problem in \ref{BPVEq} is determined by the specific structure of the function $M(x,\mu_l)$, which depends on the post-shock value function $V(x,\mu_h)$. 
This value function corresponds to the case of a constant mortality force $\mu_h$. Although the explicit expressions of the value function under a generic constant mortality force $\mu$ were already recalled in the Appendix of the main text, here we present the results tailored to the constant mortality force $\mu_h$. This formulation will facilitate the derivation of the subsequent proofs. 

\subsection{Value function under constant mortality force} \label{vfmuh} 
In this subsection, we recall from \cite{buttarazzi2024market} the explicit expression of the value function $V(x,\mu_h)$ together with the continuation and stopping sets ($\cC_h$ and $\cS_h$) for the problem with constant mortality force $\mu_t=\mu_h$ for all $t\ge 0$. We remark that \cite{buttarazzi2024market} studies the optimal stopping problem on the state space $(0,\infty)$, whereas \cite{buttarazzi2025optimal} extends the analysis to $[0,\infty)$. Hence, by applying \cite[Thm.~2.7]{buttarazzi2025optimal}, we present the continuation and stopping regions in the enlarged state space.

We distinguish cases according to the sign of $K$ and the relative magnitude of $\delta_h$ and $\beta_h$.
\begin{itemize}
    \item[\textnormal{(1)}]  Let $K > 0$.
    \begin{itemize}
    \item[i)] If $\delta_h\geq \beta_h$, then $\mathcal{C}_h=\varnothing$, $\mathcal{S}_h=[0,\infty)$ and \begin{equation} \label{V0knegdelta1bigger} V(x,\mu_h)= \delta_h(x-K).\end{equation}
    
    \item[ii)] If $\delta_h<\beta_h$, then $\mathcal{C}_h=(x^2_h,\infty)$, $\mathcal{S}_h=[0,x^2_h]$ and
    \begin{equation}  V(x,\mu_h)=\begin{cases} \label{V0knegdelta1lower} \beta_h x + \zeta^2_h x^{\gamma^-_h}, \hspace{5mm} &\text{ if } x>x^2_h,\\
    \delta_h (x-K), \hspace{5mm} &\text{ if } x\leq x^2_h,\\
    \end{cases}\end{equation}
    where $\gamma^-_h$ is recalled in \eqref{recallgamma} and $x^2_h$ and $\zeta^2_h$ are defined in Eq. (3.3) of the main text as
    \begin{align} \label{recallxdzetad}
    \begin{aligned}
    x^2_h=\frac{\delta_h K \gamma^-_h}{(\gamma^-_h-1)(\delta_h-\beta_h)}, \qquad
    \zeta^2_h=\Big(\frac{\delta_h K}{\gamma^-_h-1}\Big)^{1-\gamma^-_h}\Big(\frac{\delta_h - \beta_h}{\gamma^-_h}\Big)^{\gamma^-_h}.
    \end{aligned}
    \end{align}
    \end{itemize}
    
\item[\textnormal{(2)}]  Let $K < 0$.
    \begin{itemize}
    \item[i)] If $\delta_h\leq \beta_h$, then $\mathcal{C}_h=[0,\infty)$, $\mathcal{S}_h=\varnothing$ and \begin{equation} \label{V0kposdelta1lower} V(x,\mu_h)=\beta_h x. \end{equation} 
    
    \item[ii)] If $\delta_h>\beta_h$, then $\mathcal{C}_h=[0,x^4_h)$, $\mathcal{S}_h=[x^4_h,\infty)$ and
    \begin{equation}\label{V0kposdelta1bigger}  V(x,\mu_h)=\begin{cases} \beta_h x + \zeta^4_h x^{\gamma^+_h}, \hspace{5mm} &\text{ if } x<x^4_h,\\
    \delta_h (x-K), \hspace{5mm} &\text{ if } x\geq x^4_h,\\
    \end{cases}\end{equation}
    where $\gamma^+_h$ is recalled in \eqref{recallgamma} and $x^4_h$ and $\zeta^4_h$ are defined in Eq. (3.4) of the main text as
    \begin{align} 
    \begin{aligned}
    x^4_h=\frac{\delta_h K \gamma^+_h}{(\gamma^+_h-1)(\delta_h-\beta_h)} ,\qquad    \zeta^4_h=\Big(\frac{\delta_h K}{\gamma^+_h-1}\Big)^{1-\gamma^+_h}\Big(\frac{\delta_h - \beta_h}{\gamma^+_h}\Big)^{\gamma^+_h}.
    \end{aligned}
    \end{align}
    \end{itemize}   
    \item[\textnormal{(3)}]  Let $K = 0$. If $\delta_h < \beta_h$, then $\mathcal{C}_h=(0,\infty)$, $\mathcal{S}_h=\{0\}$ and $V(x,\mu_h)=\beta_h x$. If $\delta_h\geq \beta_h$, then $\mathcal{C}_h=\varnothing$, $\mathcal{S}_h=[0,\infty)$ and $V(x,\mu_h)= \delta_h x$.
\end{itemize}

Notice that since $\gamma^+_h>1$ and $\gamma^-_h<0$, then $x^2_h$, $\zeta^2_h$, $x^4_h$ and $\zeta^4_h$ are strictly positive if and only if $K\neq 0$ and they reduce to zero when $K=0$.

\subsection{Proofs of main results} \label{proofresultsmain}

Finally, in this subsection, we present the proof of our main results.

\begin{proof}[\textbf{Proof of Proposition 3.2 in the main text}]

First, we observe that combining the explicit expression for $V(x,\mu_h)$ in \eqref{V0knegdelta1bigger} with $M(x,\mu_l)$ in \eqref{Monejump} and using the easily verifiable equality $r_l  \delta_l\! -\! \lambda_l \delta_h=\hat{\rho} + \hat{\mu}$, we obtain 
\begin{align}
    \begin{aligned} \label{Mklowerdeltabigger}
        M(x,\mu_l)
        &=(r_l  \delta_l\! -\! \lambda_l \delta_h) K + ((\delta_l\!-\!\beta_l)(\theta\!-\!\alpha\!-\!r_l ) + \lambda_l \delta_h)x\\
        &=(\hat{\rho} + \hat{\mu}) K + M^\delta_l x,
    \end{aligned}
\end{align} 
where $M^\delta_l$ is recalled in \eqref{recallingMldeltabeta}.

First, we prove (i). Observe that $M(0,\mu_l)=(\hat{\rho} + \hat{\mu}) K<0$ and, from \eqref{Mklowerdeltabigger}, when $M^\delta_l\leq 0$ we have $M(x,\mu_l)\leq 0$ for all $x\in[0,\infty)$. Thus, from Proposition \ref{proposuffcondm}--(2.i), it follows that $\mathcal{C}_l=\varnothing$ and $\mathcal{S}_l=[0,\infty)$. In this case, clearly $W(x,\mu_l)\equiv 0$ and, using \eqref{ValuefunctionWonejump}, $V(x,\mu_l) = \delta_l (x-K)$ for $x\ge 0$.

Next, we prove (ii). From \eqref{Mklowerdeltabigger}, when $M^\delta_l> 0$, it follows that $M(x,\mu_l)$ is monotonically increasing in $x$ and diverges to $+\infty$ as $x\to \infty$. Thus, there exists $\hat{x} \in [0,\infty)$ such that $M(\hat{x},\mu_l)>0$ and from Proposition \ref{proposuffcondm}, $\mathcal{C}_l=(b,\infty)$ and $\mathcal{S}_l=[0,b]$ for some $b>0$ to be found. The unknown value function $W(x,\mu_l)$ and the unknown point $b$ must be solutions of the free-boundary problem \eqref{BPVEq}, that, in this case, reads
    \begin{align}
        (\operL-r_l ) W(x,\mu_l) &= (\hat{\rho} + \hat{\mu}) K + M^\delta_l x, \hspace{-2cm}&& x\in(b,\infty), \label{bvp2a1}\\
        W(x,\mu_l) &= 0, \hspace{-2cm}&&  x\in[0,b],\label{bvp2a2}\\
        W_x(b,\mu_l) &= 0.\label{bvp2a3}
    \end{align}
    From \eqref{bvp2a1}, one has \begin{equation} 
    W(x,\mu_l)=\zeta_1 x^{\gamma^+_l} + \zeta_2 x^{\gamma^-_l}  + \frac{ M^\delta_l }{\theta-\alpha-r_l } x -
    \frac{\hat{\rho} + \hat{\mu}}{r_l } K, \qquad x\in(b,\infty), \label{bvp2a1sol}
    \end{equation}
    where $\zeta_1$ and $\zeta_2$ are unknown constants to be found. 
    Using the sub-linear growth condition in Proposition~\ref{Vnfiniteness} (and the relation \eqref{ValuefunctionWonejump} between $V$ and $W$), it follows that $\zeta_1=0$. Hence, \eqref{bvp2a1sol} reduces to
\[
W(x,\mu_l)= \zeta_2 x^{\gamma^-_l} + \frac{ M^\delta_l }{\theta-\alpha-r_l } x - \frac{\hat{\rho} + \hat{\mu}}{r_l } K, \qquad x\in(b,\infty).
\]
    Case (ii) follows by solving \eqref{bvp2a2} and \eqref{bvp2a3}, i.e. 
    \begin{align}
    \begin{cases}
        \zeta_2 b^{\gamma^-_l} + \frac{ M^\delta_l }{\theta-\alpha-r_l } b - \frac{\hat{\rho} + \hat{\mu}}{r_l } K=0,\\
        \zeta_2 \gamma^-_l b^{\gamma^-_l-1} + \frac{ M^\delta_l }{\theta-\alpha-r_l } =0,
    \end{cases}
    \end{align}
    for two remaining unknowns $\zeta_2$ and $b$.
    
\end{proof}

\begin{proof}[\textbf{Proof of Proposition 3.3 in the main text}]
Combining $M(x,\mu_l)$ in \eqref{Monejump} and the explicit expression for $V(x,\mu_h)$ in \eqref{V0knegdelta1lower}, we obtain
\begin{align}
    \begin{aligned} \label{M1casedelta1lowerbeta1kneg}
        M(x,\mu_l) &=[(\hat{\rho}+\hat{\mu})K+ M^\delta_l x]\ind_{\{x\leq x^2_h\}}\\ 
        &\quad + (r_l  \delta_l K  +M^\beta_l x + \lambda_l\zeta^2_h x^{\gamma^-_h})\ind_{\{x>x^2_h\}},
    \end{aligned}
\end{align} 
where $M^\delta_l$ and $M^\beta_l$ are recalled in \eqref{recallingMldeltabeta}, $\zeta^2_h$ and $x^2_h$ in \eqref{recallxdzetad}.

First, we prove (i). We have $M(0,\mu_l)<0$. For $M^\beta_l \le 0$, since $\delta_h < \beta_h$, then also $M^\delta_l < 0$ (cf. \eqref{recallingMldeltabeta}). Thus, $M(x,\mu_l)$ is monotonically decreasing and it converges (as $x \to \infty$) either to $r_l \delta_l K < 0$ when $M^\beta_l = 0$, or to $-\infty$ when $M^\beta_l < 0$. This implies that $M(x,\mu_l)<0$ for all $x \in [0,\infty)$.
From Proposition \ref{proposuffcondm}, we find $\mathcal{S}_l=[0,\infty)$ and $\mathcal{C}_l=\varnothing$. In this case, we have $W(x,\mu_l)\equiv 0$. 

Next, we prove (ii).
From \eqref{Mklowerdeltabigger}, when $M^\beta_l> 0$, it follows that $M(x,\mu_l)$ is monotonically increasing in $x$ and diverges to $+\infty$ as $x\to \infty$. Thus, there exists $\hat{x} \in [0,\infty)$ such that $M(\hat{x},\mu_l)>0$. 
From Proposition \ref{proposuffcondm}, $\mathcal{C}_l=(b,\infty)$ and $\mathcal{S}_l=[0,b]$ for some $b>0$ to be found.
Now, we must solve the free boundary problem \eqref{BPVEq}. 
From \eqref{M1casedelta1lowerbeta1kneg}, letting 
\begin{align} \begin{aligned} \label{Mcheck}\check{M}(x,\mu_l)\coloneqq(\hat{\rho}+\hat{\mu})K+ M^\delta_l x,\end{aligned}\end{align}  
and
\begin{align} \begin{aligned} \hat{M}(x,\mu_l)\coloneqq r_l  \delta_l K  +M^\beta_l x + \lambda_l\zeta^2_h x^{\gamma^-_h}, \end{aligned}\end{align}  
we have $M(x,\mu_l)=\check{M}(x,\mu_l) \ind_{\{x\leq x^2_h\}}+ \hat{M}(x,\mu_l) \ind_{\{x>x^2_h\}}$.
We address the solution to the free-boundary problem in \eqref{BPVEq} in two steps. First, we analyse the case where the boundary point $b$ satisfies $b > x^2_h$. Then, we consider the case $b \leq x^2_h$.

\begin{itemize}
    \item[(ii.1)] For $b>x^2_h$, the free boundary problem \eqref{BPVEq} becomes

    \begin{align}
        (\operL-r_l ) W(x,\mu_l) &= \hat{M}(x,\mu_l),  \hspace{-4cm} &&x\in (b,\infty),\label{bvp2bb1}\\
        W(x,\mu_l) &= 0, \hspace{-4cm}  &&x\in  [0,b],\label{bvp2bb2}\\
        W_x(b,\mu_l)&= 0.\label{bvp2bb3}
    \end{align}
    We now solve an ordinary differential equation \eqref{bvp2bb1} using the well-known \textit{variation of parameter} method (cf. \cite{agarwal2008ordinary}). To this end, we introduce the Wronskian $\mathcal{W}$ defined as
    \begin{align}
        \begin{aligned}
        \mathcal{W}(x,\mu_l)&\coloneqq (\gamma^-_l - \gamma^+_l) x^{\gamma^+_l + \gamma^-_l -1},
        \end{aligned}
    \end{align}  
    where $\gamma^{\pm}_l$ are recalled in \eqref{recallgamma}.
    Thus, from \eqref{bvp2bb1} and using \textit{variation of parameter} method, it yields
    \begin{align}
        \begin{aligned}
            \label{Odebvp2bbb3} W(x,\mu_l)
            &=\zeta_1 x^{\gamma^+_l} + \zeta_2 x^{\gamma^-_l} + W^p(x),
        \end{aligned}
    \end{align}
    where $\zeta_1$ and $\zeta_2$ are undetermined constants and 
    \begin{align} \begin{aligned} \label{w1particolarecasoxdminorexf} W^p(x)= -x^{\gamma^+_l} \int \frac{2x^{\gamma^-_l}\hat{M}(x,\mu_l)}{\sigma x^2 \mathcal{W}(x,\mu_l)}dx+x^{\gamma^-_l} \int \frac{2x^{\gamma^+_l}\hat{M}(x,\mu_l)}{\sigma x^2 \mathcal{W}(x,\mu_l)}dx.  \end{aligned}\end{align} 
    From \eqref{w1particolarecasoxdminorexf}, one has
    \begin{align}
        \begin{aligned}\label{w1particolarecasoxdminorexf2}
        W^p(x)&=\frac{2 r_l \delta_lK}{\sigma^2 \gamma^+_l \gamma^-_l}+\frac{2 M^\beta_l}{\sigma^2 (1-\gamma^+_l)(1-\gamma^-_l)} x \\
        &\quad + \frac{2 \lambda_l \zeta^2_h}{\sigma^2 (\gamma^-_h-\gamma^+_l)(\gamma^-_h-\gamma^-_l)} x^{\gamma^-_h}.
        \end{aligned}
    \end{align}
    
    Using the easily verifiable equalities 
    \begin{align} \begin{aligned} &\sigma^2 \gamma^+_l \gamma^-_l= -2 r_l,\\
    &\sigma^2 (1-\gamma^+_l)(1-\gamma^-_l)=2(\theta-\alpha-r_l),\\
    &\sigma^2 (\gamma^-_h-\gamma^+_l)(\gamma^-_h-\gamma^-_l)=2 (\Delta-\lambda_l),
        \end{aligned}
    \end{align}
    we rewrite \eqref{w1particolarecasoxdminorexf2} as
    \begin{align}
        \begin{aligned}
        W^p(x)&=-\delta_lK+\frac{M^\beta_l}{\theta-\alpha-r_l} x + \frac{ \lambda_l \zeta^2_h}{\Delta-\lambda_l} x^{\gamma^-_h},  
        \end{aligned}
    \end{align}
    and combining the latter with \eqref{Odebvp2bbb3}, we find
    \begin{align}
        \begin{aligned} \label{bvp2bb1sol}
           W(x,\mu_l)=\zeta_1 x^{\gamma^+_l} + \zeta_2 x^{\gamma^-_l} -\delta_lK+\frac{M^\beta_l}{\theta-\alpha-r_l} x + \frac{ \lambda_l \zeta^2_h}{\Delta-\lambda_l} x^{\gamma^-_h}.
        \end{aligned}
    \end{align}
    Using Proposition \ref{Vnfiniteness}, then we find $\zeta_1=0$. 
    Finally, case (ii.1) follows by solving \eqref{bvp2bb2}-\eqref{bvp2bb3}, for the two unknowns $\zeta_2$ and $b$.
    
    \item[(ii.2)] For $b\leq x^2_h$ the free-boundary problem \eqref{BPVEq} becomes
    \begin{align}
        (\operL-r_l ) W(x,\mu_l) &= \check{M}(x,\mu_l),  \hspace{-3cm} &&x\in  (b,x^2_h],\label{bvp2bbb0}\\
        (\operL-r_l ) W(x,\mu_l) &= \hat{M}(x,\mu_l), \hspace{-3cm} &&x\in  (x^2_h,\infty), \label{bvp2bbb1}\\
        W(x,\mu_l) &= 0, \hspace{-3cm} &&x\in  [0,b],\label{bvp2bbb2}\\
        W_x(b,\mu_l)&= 0. \label{bvp2bbb3}
    \end{align}

    The ODE \eqref{bvp2bbb0} is the same as \eqref{bvp2a1}. Thus, for $x\in (b,x^2_h]$
    \begin{equation} \label{ContW0}
        W(x,\mu_l)=\zeta_1 x^{\gamma^+_l} + \zeta_2 x^{\gamma^-_l}  + \frac{ M^\delta_l }{\theta-\alpha-r_l } x -
        \frac{\hat{\rho} + \hat{\mu}}{r_l } K,
        \end{equation}
    with $\zeta_1$, $\zeta_2$ undetermined constants.
    Similarly, \eqref{bvp2bbb1} is the same as \eqref{bvp2bb1}. Hence, in $(x^2_h,\infty)$,
    \begin{align}
        \begin{aligned} \label{ContW1}
           W(x,\mu_l)=\zeta_3 x^{\gamma^+_l} + \zeta_4 x^{\gamma^-_l} -\delta_lK+\frac{M^\beta_l}{\theta-\alpha-r_l} x + \frac{ \lambda_l \zeta^2_h}{\Delta-\lambda_l} x^{\gamma^-_h},
        \end{aligned}
    \end{align}
    with $\zeta_3$, $\zeta_4$ undetermined constants. Using Proposition \ref{Vnfiniteness}, then we get $\zeta_3=0$.
    We now require that $W(x,\mu_l)$ is continuously differentiable, see Proposition \ref{regularityvf}.
    Imposing the continuity of $W(x,\mu_l)$ in $x^2_h$ and using \eqref{ContW0} and \eqref{ContW1}, we get
    \begin{align}
        \begin{aligned}
         &\zeta_1 {(x^2_h)}^{\gamma^+_l} + (\zeta_2-\zeta_4){(x^2_h)}^{\gamma^-_l} \\
         &\quad + \bigg(
    \delta_l - \frac{\hat{\rho} + \hat{\mu}}{r_l } \bigg)K+\frac{ (M^\delta_l-M^\beta_l)x^2_h }{\theta-\alpha-r_l }  - \frac{\lambda_l \zeta^2_h{(x^2_h)}^{\gamma^-_h} }{\Delta-\lambda_l}=0.   
        \end{aligned}
    \end{align}
    Using the following verifiable equalities
    \begin{align} \begin{aligned} &\delta_l - \frac{\hat{\rho} + \hat{\mu}}{r_l }= \frac{\lambda_l \delta_h}{r_l}, \\
    &(M^\delta_l-M^\beta_l)x^2_h=\frac{\gamma^-_h \lambda_l\delta_hK}{\gamma^-_h-1}, \\
    &\zeta^2_h{(x^2_h)}^{\gamma^-_h}=\frac{\delta_h K}{\gamma^-_h-1}, \end{aligned}\end{align} 
    we obtain the expression
    \begin{align} \begin{aligned} \label{zeta4} \zeta_4=\zeta_1 (x^2_h)^{\gamma^+_l - \gamma^-_l} + \zeta_2 + \pi^2_l (x^2_h)^{-\gamma^-_l} \end{aligned}\end{align} 
    with 
    \begin{align}
        \pi^2_l=\bigg[\frac{1}{r_l} + \frac{\gamma^-_h}{(\theta-\alpha-r_l)(\gamma^-_h-1)}-\frac{1}{(\Delta-\lambda_l)(\gamma^-_h-1)} \bigg]\lambda_l \delta_h K.
    \end{align}
    Substituting $\zeta_3=0$ and the expression for $\zeta_4$ from \eqref{zeta4} into \eqref{ContW1}, and combining with \eqref{bvp2bbb2} and \eqref{ContW0}, we find \begin{align} \begin{aligned}
    W(x,\mu_l) =
    \begin{cases}
        0, & x \in [0,b],\\
        \\
        \zeta_1 x^{\gamma^+_l} + \zeta_2 x^{\gamma^-_l} + \frac{M^\delta_l}{\theta - \alpha - r_l} x - \frac{\hat{\rho} + \hat{\mu}}{r_l} K, & x \in (b, x^2_h], \\
        \\
        \begin{aligned}
            &\Big( \zeta_1 (x^2_h)^{\gamma^+_l - \gamma^-_l} + \zeta_2 + \pi^2_l (x^2_h)^{-\gamma^-_l} \Big) x^{\gamma^-_l} \\
            & - \delta_lK + \frac{M^\beta_l}{\theta - \alpha - r_l} x + \frac{\lambda_l \zeta^2_h}{\Delta-\lambda_l} x^{\gamma^-_h},
        \end{aligned} & x \in (x^2_h, \infty).
    \end{cases}
    \end{aligned}\end{align} 
    We now impose the continuity of $W_x(x,\mu_l)$ at $x^2_h$, yielding
    \begin{align} \begin{aligned} (\gamma^+_l-\gamma^-_l)\zeta_1 (x^2_h)^{\gamma^+_l-1} =  \frac{\pi^2_l}{x^2_h} \gamma^-_l +  \frac{ \lambda_l}{\Delta-\lambda_l} \gamma^-_h \zeta^2_h {(x^2_h)}^{\gamma^-_h-1} - \frac{M^\delta_l - M^\beta_l}{\theta-\alpha-r_l}. \end{aligned}\end{align} 
    Using that $\gamma^-_h \zeta^2_h {(x^2_h)}^{\gamma^-_h-1} = \delta_h-\beta_h$ and
    \begin{align} \begin{aligned} \frac{\pi^2_l}{x^2_h}=\bigg[ \frac{(\gamma^-_h-1)}{r_l \gamma^-_h} + \frac{1}{\theta-\alpha-r_l} - \frac{1}{\gamma^-_h(\Delta-\lambda_l)} \bigg]\lambda_l(\delta_h-\beta_h),\end{aligned}\end{align}  
    we find 
    $\zeta_1=\varpi^2_l (x^2_h)^{1-\gamma^+_l}$
    with 
    \begin{align}
        \varpi^2_l\bigg[\frac{\gamma^-_l-1}{\theta-\alpha-r_l}+\frac{\gamma^-_l(\gamma^-_h-1)}{r_l\gamma^-_h}+\frac{\gamma^-_h-\gamma^-_l}{\gamma^-_h(\Delta-\lambda_l)}\bigg]\frac{\lambda_l(\delta_h-\beta_h)}{\gamma^+_l-\gamma^-_l}.
    \end{align}
    Putting all together,
    \begin{align} \begin{aligned} W(x,\mu_l)=
    \begin{cases}
    0, & x \in [0,b], \\
    \\
    \varpi^2_l (x^2_h)^{1-\gamma^+_l} x^{\gamma^+_l} + \zeta_2 x^{\gamma^-_l}  + \frac{ M^\delta_l }{\theta-\alpha-r_l } x -
        \frac{\hat{\rho} + \hat{\mu}}{r_l } K, & x \in (b, x^2_h], \\
    \\
    \begin{aligned} 
        & (\varpi^2_l (x^2_h)^{1-\gamma^-_l}  +\pi^2_l (x^2_h)^{-\gamma^-_l})x^{\gamma^-_l} + \zeta_2 x^{\gamma^-_l} \\
        &\quad -\delta_lK+\frac{M^\beta_l}{\theta-\alpha-r_l} x + \frac{ \lambda_l \zeta^2_h}{\Delta-\lambda_l} x^{\gamma^-_h},
        \end{aligned} & x \in (x^2_h, \infty).
    \end{cases}
    \end{aligned}\end{align} 
    Solving \eqref{bvp2bbb2}-\eqref{bvp2bbb3} for the two unknown $\zeta_2$ and $b$ completes the proof.
\end{itemize}
\end{proof}

\begin{proof}[\textbf{Proof of Proposition 3.4 in the main text}]
First observe that combining \eqref{Monejump} and \eqref{V0kposdelta1lower},
\begin{align}
    \begin{aligned} \label{mincase1}
        M(x,\mu_l)
        &=r_l  \delta_l K + ((\delta_l-\beta_l)(\theta-\alpha-r_l ) + \lambda_l \beta_h)x\\
        &=r_l  \delta_l K + M^\beta_l x,
    \end{aligned}
\end{align}
with $M^\beta_l$ recalled in \eqref{recallingMldeltabeta}. 
Now, we start proving (i). For $M^\beta_l<0$, it follows that $M(x,\mu_l)$ diverges to $-\infty$ as $x\to\infty$.
By Proposition \ref{proposuffcondm}, we conclude that the continuation and stopping regions are given by $\mathcal{C}_l=[0,b)$ and $\mathcal{S}_l=[b,\infty)$ for some $b>0$. We now must solve the free-boundary problem \eqref{BPVEq}, with $M(x,\mu_l)$ given by \eqref{mincase1}, to characterize the boundary point $b$ and the value function $W(x,\mu_l)$. Specifically, we have
    \begin{align}
        (\operL-r_l ) W(x,\mu_l) &= r_l  \delta_l K + M^\beta_l x, \hspace{-2cm}&& x\in(0,b), \label{bvp2c1}\\
        W(x,\mu_l) &= 0, \hspace{-2cm}&&  x\in[b,\infty),\label{bvp2c2}\\
        W_x(b,\mu_l) &= 0.\label{bvp2c3}
    \end{align}
    
    From \eqref{bvp2c1} one can obtain \begin{equation}
    W(x,\mu_l)=\zeta_1 x^{\gamma_l^+} + \zeta_2 x^{\gamma_l^-}  + \frac{ M^\beta_l }{\theta-\alpha-r_l } x - \delta_l  K, \qquad x\in \mathcal{C}, \label{bvp2b1sol}
    \end{equation}
    with $\zeta_1$, $\zeta_2$ constants to be found. Using the sub-linear growth of the value function, we get $\zeta_2=0$. Case (i) follows by solving \eqref{bvp2c2} and \eqref{bvp2c3} for two remaining unknowns $\zeta_1$ and $b$.

We now prove (ii). Since $M(0,\mu_l)>0$ and $M^\beta_l\geq 0$, then $M(x,\mu_l)>0$ for all $x\in [0,\infty)$. It follows from Proposition \ref{proposuffcondm} that $\mathcal{C}_l=[0,\infty)$ and $\mathcal{S}_l=\varnothing$. In this case, we find \begin{align}
    W(x,\mu_l)&= \E_x \bigg[\int_0^\infty e^{-r_lt} M(X_t,\mu_l) \ud t \bigg]\\
    &= \int_0^\infty e^{-r_l t} r_l  \delta_l K  \ud t + \int_0^\infty e^{-r_l t} M^\beta_l \E_x[X_t] \ud t\\
    &= \delta_l K + \frac{M^\beta_l}{r_l + \alpha - \theta} x.
\end{align}
Recalling the relation \eqref{ValuefunctionWonejump} between $V$ and $W$, the proof is complete.
\end{proof}

\begin{proof}[\textbf{Proof of Proposition 3.5 in the main text}]

Combining $M(x,\mu_l)$ in \eqref{Monejump} and the explicit expression for $V(x,\mu_h)$ in \eqref{V0kposdelta1bigger}, and using $r_l  \delta_l - \lambda_l \delta_h=\hat{\rho} + \hat{\mu}$, we get  \begin{align}
    \begin{aligned} \label{Mdmagb}
        M(x,\mu_l)
        &=r_l  \delta_l K + (\delta_l-\beta_l)(\theta-\alpha-r_l )x + \hspace{1cm}\\
        &\quad +\lambda_l\delta_h (x-K)\ind_{\{x\geq x^4_h\}} + \lambda_l\big(\beta_h x + \zeta^4_h x^{\gamma^+_h}\big)\ind_{\{x< x^4_h\}}\\
        &=\ind_{\{x< x^4_h\}}  (r_l  \delta_l K + M^\beta_l x + \lambda_l\zeta^4_h x^{\gamma^+_h})\\
        &\quad + \ind_{\{x\geq x^4_h\}} ( (\hat{\rho}+\hat{\mu})K+ M^\delta_l x).
    \end{aligned}
\end{align}
We begin by proving (i). When $M^\delta_l<0$ it follows that $M(x,\mu_l)$ goes to $-\infty$ as $x\to\infty$ and by proposition \ref{proposuffcondm}, $\mathcal{C}_l=[0,b)$ and $\mathcal{S}_l=[b,\infty)$ for some $b>0$. We now must solve the free-boundary problem \eqref{BPVEq}, with $M(x,\mu_l)$ given by \eqref{Mdmagb}. 
Letting 
\begin{align}
\tilde{M}(x)\coloneqq r_l  \delta_l K + M^\beta_l x + \lambda_l \zeta^4_h x^{\gamma^+_h}.
\end{align}
we write
\[M(x,\mu_l)=\Tilde{M}(x) \ind_{\{x< x^4_h\}} + \check{M}(x,\mu_l) \ind_{\{x\geq x^4_h\}}, \] where $\check{M}(x,\mu_l)$ is defined in \eqref{Mcheck}.
Analogously to the proof of Proposition 3.3 of the main text, we split the analysis into two cases. 

\begin{itemize}
    \item[(i.1)] For $b<x^4_h$, the free-boundary problem \eqref{BPVEq} becomes
    \begin{align}
    (\operL-r_l ) W(x,\mu_l) &= \tilde{M}(x), \hspace{-2cm} &&x\in  (0,b), \label{app2bvp2bb1}\\
    W(x,\mu_l) &= 0, \hspace{-2cm} &&x\in  [b,\infty),\label{app2bvp2bb2} \\
    W_x(b,\mu_l)&= 0. \hspace{-2cm} &&  \label{app2bvp2bb3}&
\end{align}
Since the structure of \eqref{app2bvp2bb1} closely mirrors that of \eqref{bvp2bb1}, we solve it analogously and, omitting intermediate steps, we find
    \begin{align}
        \begin{aligned} \label{app2bvp2bb1sol}
           W(x,\mu_l)=\zeta_1 x^{\gamma^+_l} + \zeta_2 x^{\gamma^-_l} -\delta_lK+\frac{M^\beta_l}{\theta-\alpha-r_l} x + \frac{ \lambda_l \zeta^4_h}{\Delta-\lambda_l} x^{\gamma^+_h}, & \ \ \ x\in  (0,b),
        \end{aligned}
    \end{align}
    where $\zeta_1$ and $\zeta_2$ are undetermined constants. Using Proposition \ref{Vnfiniteness}, we get $\zeta_2=0$. Solving \eqref{app2bvp2bb2}-\eqref{app2bvp2bb3} for the unknowns $b$ and $\zeta_1$ and using the relation between $V$ and $W$ in \eqref{ValuefunctionWonejump} yield the explicit expressions stated in item (i.1) of the proposition.
   
    \item[(i.2)] For $b \geq x^4_h$, the free-boundary problem \eqref{BPVEq} becomes
        \begin{align}
        (\operL-r_l ) W(x,\mu_l) &= \tilde{M}(x),  \hspace{-3cm} &&x\in  (0,x^4_h),\label{app2bvp2bbb0}\\
        (\operL-r_l ) W(x,\mu_l) &= \check{M}(x,\mu_l), \hspace{-3cm} &&x\in (x^4_h,b), \label{app2bvp2bbb1}\\
        W(x,\mu_l) &= 0, \hspace{-3cm} &&x\in  [b,\infty),\label{app2bvp2bbb2}\\
        W_x(b,\mu_l)&= 0. \hspace{-3cm} &&\label{app2bvp2bbb3}
    \end{align}

    Analogously to \eqref{app2bvp2bb1}, from \eqref{app2bvp2bbb0}, we find 
        \begin{align}
        \begin{aligned}
           W(x,\mu_l)=\zeta_1 x^{\gamma^+_l} + \zeta_2 x^{\gamma^-_l} -\delta_lK+\frac{M^\beta_l}{\theta-\alpha-r_l} x + \frac{ \lambda_l \zeta^4_h}{\Delta-\lambda_l} x^{\gamma^+_h}, & \qquad x\in   (0,x^4_h)
        \end{aligned}
    \end{align} 
    with $\zeta_1$, $\zeta_2$ undetermined constants.
    The ODE \eqref{app2bvp2bbb1} is the same as \eqref{bvp2bbb0}. Thus, 
    \begin{equation} 
        W(x,\mu_l)=\zeta_3 x^{\gamma^+_l} + \zeta_4 x^{\gamma^-_l}  + \frac{ M^\delta_l }{\theta-\alpha-r_l } x -
        \frac{\hat{\rho} + \hat{\mu}}{r_l } K, \qquad x\in (x^4_h,b)
        \end{equation}
    with $\zeta_3$, $\zeta_4$ undetermined constants.
    Using Proposition \ref{Vnfiniteness}, we get $\zeta_2=0$. We now impose that $W(x,\mu_l)$ is continuously differentiable, see Proposition \ref{regularityvf}.
    Imposing the continuity of $W(x,\mu_l)$ in $x^4_h$, we find
    \begin{align}
        \begin{aligned}
        &\zeta_1 {(x^4_h)}^{\gamma^+_l} = \zeta_3 {(x^4_h)}^{\gamma^+_l} + \zeta_4{(x^4_h)}^{\gamma^-_l} \\
        &\quad + \bigg(\delta_l - \frac{\hat{\rho} + \hat{\mu}}{r_l } \bigg)K+\frac{ (M^\delta_l-M^\beta_l)x^4_h }{\theta-\alpha-r_l }  - \frac{\lambda_l \zeta^4_h{(x^4_h)}^{\gamma^+_h} }{\Delta-\lambda_l}.   
        \end{aligned}
    \end{align}
    Using the following verifiable equalities 
    \begin{align} \begin{aligned} &(M^\delta_l-M^\beta_l)x^4_h=\frac{\gamma^+_h \lambda_l\delta_hK}{\gamma^+_h-1}, \qquad \text{and} \qquad \zeta^4_h{(x^4_h)}^{\gamma^+_h}=\frac{\delta_h K}{\gamma^+_h-1}, \end{aligned}\end{align} 
    it yields
    \begin{align} \begin{aligned} \zeta_1=\zeta_3 + \zeta_4 (x^4_h)^{\gamma^-_l - \gamma^+_l} + \pi^4_l (x^4_h)^{-\gamma^+_l} \end{aligned}\end{align} 
    with 
    \begin{align}
        \pi^4_l=\Big[\frac{1}{r_l} + \frac{\gamma^+_h}{(\theta-\alpha-r_l)(\gamma^+_h-1)}-\frac{1}{(\Delta-\lambda_l)(\gamma^+_h-1)} \Big]\lambda_l \delta_h K.
    \end{align}
    We can rewrite the value function as \begin{align} \begin{aligned}
    W(x,\mu_l) =
    \begin{cases}\begin{aligned}
    &\bigg(\zeta_3  + \zeta_4 (x^4_h)^{\gamma^-_l - \gamma^+_l} + \pi^4_l (x^4_h)^{-\gamma^+_l}\bigg)x^{ \gamma^+_l} \\
    & \quad -\delta_lK+\frac{M^\beta_l}{\theta-\alpha-r_l} x + \frac{ \lambda_l \zeta^4_h}{\Delta-\lambda_l} x^{\gamma^+_h},
    \end{aligned} & x \in (0, x^4_h], \\
        \\
        \zeta_3 x^{\gamma^+_l} + \zeta_4 x^{\gamma^-_l}  + \frac{ M^\delta_l }{\theta-\alpha-r_l } x -
        \frac{\hat{\rho} + \hat{\mu}}{r_l } K,  & x \in (x^4_h, b),\\
        \\
        0, & x \in [b,\infty).
    \end{cases}
    \end{aligned}\end{align} 
    Imposing the continuity of $W_x(x,\mu_l)$ in $x^4_h$, we find
    \begin{align} \begin{aligned} (\gamma^+_l-\gamma^-_l)\zeta_4 (x^4_h)^{\gamma^-_l-1} = \frac{M^\delta_l - M^\beta_l}{\theta-\alpha-r_l} - \frac{\pi^4_l\gamma^+_l}{x^4_h}  -  \frac{ \lambda_l \gamma^+_h \zeta^4_h {(x^4_h)}^{\gamma^+_h-1}}{\Delta-\lambda_l} . \end{aligned}\end{align} 
    Observing that
    \begin{align} \begin{aligned} &\frac{\pi^4_l}{x^4_h}=\bigg[ \frac{(\gamma^+_h-1)}{r_l \gamma^+_h} + \frac{1}{\theta-\alpha-r_l} - \frac{1}{\gamma^+_h(\Delta-\lambda_l)} \bigg]\lambda_l(\delta_h-\beta_h),\\
    &\gamma^+_h \zeta^4_h {(x^4_h)}^{\gamma^+_h-1} = \delta_h-\beta_h,\end{aligned}\end{align}  
    it follows that
    $\zeta_4=\varpi^4_l (x^4_h)^{1-\gamma^-_l}$
    with 
    \[\varpi^4_l=\Big[\frac{(1-\gamma^+_l)}{\theta-\alpha-r_l}-\frac{\gamma^+_l(\gamma^+_h-1)}{r_l\gamma^+_h}+\frac{\gamma^+_l-\gamma^+_h}{\gamma^+_h(\Delta-\lambda_l)}\Big]\frac{\lambda_l(\delta_h-\beta_h)}{\gamma^+_l-\gamma^-_l},\] yielding \begin{align} \begin{aligned}
    W(x,\mu_l) =
    \begin{cases}\begin{aligned}
    &\bigg(\zeta_3  + \varpi^4_l (x^4_h)^{1 - \gamma^+_l} + \pi^4_l (x^4_h)^{-\gamma^+_l}\bigg)x^{ \gamma^+_l} \\
    & \quad -\delta_lK+\frac{M^\beta_l}{\theta-\alpha-r_l} x + \frac{ \lambda_l \zeta^4_h}{\Delta-\lambda_l} x^{\gamma^+_h},
    \end{aligned} & x \in (0, x^4_h], \\
        \\
        \zeta_3 x^{\gamma^+_l} + \varpi^4_l (x^4_h)^{1-\gamma^-_l} x^{\gamma^-_l}  + \frac{ M^\delta_l }{\theta-\alpha-r_l } x -
        \frac{\hat{\rho} + \hat{\mu}}{r_l } K,  & x \in (x^4_h, b),\\
        \\
        0, & x \in [b,\infty).
    \end{cases}
    \end{aligned}\end{align} 
    Finally, solving \eqref{app2bvp2bbb2}-\eqref{app2bvp2bbb3} for the unknowns $b$ and $\zeta_3$ yields the explicit expressions stated in item (i.2) of the proposition.
    \end{itemize}

We now prove (ii). From \eqref{Mdmagb}, we find
\begin{align}
    &M_x(x,\mu_l)=( M^\beta_l  + \lambda_l\zeta^4_h \gamma^+_h  x^{\gamma^+_h -1})\ind_{\{x< x^4_h\}} +  M^\delta_l \ind_{\{x\geq x^4_h\}}\\
    &M_{xx}(x,\mu_l)= \lambda_l\zeta^4_h \gamma^+_h (\gamma^+_h-1) x^{\gamma^+_h -2}\ind_{\{x< x^4_h\}}.
\end{align}
    Thus, $M(x,\mu_l)$ is convex in $[0,x^4_h)$ and linear in $[x^4_h,\infty)$.
    Letting \begin{align} \begin{aligned} \label{xm} x_m\coloneqq\bigg(-\frac{M^\beta_l}{\lambda_l \zeta^4_h \gamma^+_h}\bigg)^{\frac{1}{\gamma^+_h-1}}, \end{aligned}\end{align} we distinguish three cases.
    \begin{itemize}
        \item Case $x_m\leq 0$. Here, $M(x,\mu_l)$ has a minimum in $x=0$. From \eqref{Mdmagb}, we have $M(0,\mu_l)>0$, and thus $M(x,\mu_l)>0$ for all $x\ge0$.
        
        \item Case $x_m\in(0,x^4_h)$. $M(x,\mu_l)$ has a minimum in $x=x_m$ and 
        \begin{align} \begin{aligned} \label{Mxm} M(x_m,\mu_l)=r_l  \delta_l K +(- M^\beta_l)^{\frac{\gamma^+_h}{\gamma^+_h-1}} (\lambda_l\zeta^4_h)^{\frac{1}{1-\gamma^+_h}} \bigg( (\gamma^+_h)^{\frac{\gamma^+_h}{1-\gamma^+_h}}-(\gamma^+_h)^{\frac{1}{1-\gamma^+_h}} \bigg). \end{aligned}\end{align} 
        This expression is always positive, implying $M(x,\mu_l)>0$ for all $x\ge0$.

        \item Case $x_m\geq x^4_h$. Here, for $x\ge0$ \begin{align} \begin{aligned} M(x,\mu_l)\ge M(x^4_h,\mu_l)>0.\end{aligned}\end{align} 
    \end{itemize}
     
    In all cases, we conclude that $M(x,\mu_l)>0$ for all $x\in[0,\infty)$. By Proposition \ref{proposuffcondm}, $\mathcal{C}_l=[0,\infty)$ and $\mathcal{S}_l=\varnothing$. In this case, the value function is 
    \begin{align}
    \begin{aligned} \label{startingw}
    W(x,\mu_l)&=\E_x \bigg[\int_0^\infty \e^{-r_l t}  M(X_t,\mu_l)  \ud t \bigg],
    \end{aligned}
\end{align}
and, in the remainder of the proof, we compute it explicitly. First, by Fubini's Theorem and using \eqref{Mdmagb}, it yields
\begin{align}
    \begin{aligned} \label{startingw}
    W(x,\mu_l)&=\int_0^\infty \e^{-r_l t}\E_x[  M(X_t^x,\mu_l)]  \ud t \\
    &=\int_0^\infty \e^{-r_l t} r_l  \delta_l K  \ud t  + \int_0^\infty \e^{-r_l t} (\delta_l-\beta_l)(\theta-\alpha-r_l )\E[ X_t] \ud t  \\
        &\quad + \int_0^\infty \e^{-r_l t}\lambda_l \E_x [ V(X_t,\mu_h)]  \ud t \\
        &=\beta_lx - \delta_l(x-K) + \int_0^\infty \e^{-r_l t}\lambda_l \E_x [ V(X_t,\mu_h)]  \ud t .
    \end{aligned}
\end{align}

We now compute explicitly $\E_x[V(X_t,\mu_h)]$. From \eqref{V0kposdelta1bigger}, we get 
\begin{align}
        \begin{aligned} \label{valoreA}
                \E [V(X_t^x,\mu_h)]&= \beta_h \E[X_t \ind_{\{X_t<x^4_h\}}] + \zeta^4_h \E[X_t^{\gamma^+_h} \ind_{\{X_t<x^4_h\}}]\\
                &+ \delta_h \E[X_t \ind_{\{X_t\geq x^4_h\}}]- \delta_h K \E[\ind_{\{X_t\geq x^4_h\}}].
        \end{aligned}
    \end{align}
Let \begin{align}
    \begin{aligned}
        d_1(x,t)\coloneqq\frac{ln\big(\frac{x}{x^4_h}\big)+(\theta-\alpha+\frac{\sigma^2}{2})t}{\sigma\sqrt{t}}, \qquad
        d_2(x,t)\coloneqq\frac{ln\big(\frac{x}{x^4_h}\big)+(\theta-\alpha-\frac{\sigma^2}{2})t}{\sigma\sqrt{t}},
    \end{aligned}
\end{align} and $\Phi$ be the cumulative distribution function of a random variable $Z$ with a standard normal distribution.
Observing that $X^x_t \ge x^4_h \iff Z\ge -d_2(x,t)$, we immediately get
    \begin{align}
        \begin{aligned} \label{valoreA1}
                \E[\ind_{\{X_t\geq x^4_h\}}]=\P(Z\geq-d_2(x,t))=\Phi(d_2(x,t)).
        \end{aligned}
    \end{align}
    Then, we can compute
    \begin{align}
        \begin{aligned}\label{valoreA2}
                \E[X_t \ind_{\{X_t<x^4_h\}}]&=\int_\R \ind_{\{z<-d_2(x,t)\}}  \ x \e^{(\theta-\alpha-\frac{\sigma^2}{2})t+\sigma z \sqrt{t}} \ \  \frac{\e^{-\frac{z^2}{2}}}{\sqrt{2\pi}}  \ \  dz\\
                &=\int_{-\infty}^{-d_2(x,t)} x \e^{(\theta-\alpha)t}  \  \frac{\e^{-\frac{1}{2}(z-\sigma\sqrt{t})^2}}{\sqrt{2\pi}}  \ \  dz\\
                &=x \e^{(\theta-\alpha)t} \int_{-\infty}^{-d_2(x,t)-\sigma\sqrt{t}} \frac{\e^{-\frac{1}{2}u^2}}{\sqrt{2\pi}}  \ \  du\\
                &=x \e^{(\theta-\alpha)t} \Phi(-d_2(x,t)-\sigma\sqrt{t})\\
                &=x \e^{(\theta-\alpha)t} \Phi(-d_1(x,t)).
        \end{aligned}
    \end{align}
    Similarly,
    \begin{align}
        \begin{aligned} \label{valoreA3}
                \E[X_t \ind_{\{X_t\geq x^4_h\}}]&=x \e^{(\theta-\alpha)t} \Phi(d_1(x,t)).
        \end{aligned}
    \end{align}
    Moreover, we have
    \begin{align}
        \begin{aligned} \label{valoreA4}
                \E[X_t^{\gamma^+_h} \ind_{\{X_t<x^4_h\}}]&=\int_{-\infty}^{-d_2(x,t)} x^{\gamma^+_h} \e^{\gamma^+_h(\theta-\alpha-\frac{\sigma^2}{2})t+\gamma^+_h \sigma z \sqrt{t}} \ \  \frac{\e^{-\frac{z^2}{2}}}{\sqrt{2\pi}}  \ \  dz\\
                &=x^{\gamma^+_h} \e^{\gamma^+_h(\theta-\alpha)t+\frac{1}{2}\sigma^2 \gamma^+_h(\gamma^+_h-1)t} \int_{-\infty}^{-d_2(x,t)} \frac{\e^{-\frac{1}{2}(z-\sigma \gamma^+_h \sqrt{t})^2}}{\sqrt{2\pi}}  \ \  dz\\
                &=x^{\gamma^+_h} \e^{\gamma^+_h(\theta-\alpha)t+\frac{1}{2}\sigma^2 \gamma^+_h(\gamma^+_h-1)t} \Phi(-d_2(x,t)-\sigma\gamma^+_h\sqrt{t}),
        \end{aligned}
    \end{align}
    and using that 
    \begin{equation} \frac{1}{2}\sigma^2 \gamma^+_h(\gamma^+_h-1) = r_h  - \gamma^+_h(\theta-\alpha)\end{equation}
    we get 
    \begin{align}
        \begin{aligned} \label{valoreA5}
                \E[X_t^{\gamma^+_h} \ind_{\{X_t<x^4_h\}}]&=x^{\gamma^+_h} \e^{r_h t} \Phi(-d_2(x,t)-\sigma\gamma^+_h\sqrt{t}).
        \end{aligned}
    \end{align}
    Finally, combining \eqref{valoreA} with \eqref{valoreA1}, \eqref{valoreA2}, \eqref{valoreA3} and \eqref{valoreA5}, we get
        \begin{align}
        \begin{aligned} \label{valoreA6}
                \E [ V(X_t,\mu_h)]&= x\e^{(\theta-\alpha)t}[\beta_h\Phi(-d_1(x,t))+\delta_h\Phi(d_1(x,t))]-\delta_h K \Phi(d_2(x,t))\\
                &\ \ \  + \e^{r_h  t}  \zeta^4_h x^{\gamma^+_h} \Phi(-d_2(x,t)-\sigma\gamma^+_h\sqrt{t}) 
        \end{aligned}
    \end{align}
Thus, plugging \eqref{valoreA6} into \eqref{startingw}, we get
\begin{align}
\begin{aligned}
\label{defWvaralpha}
    W(x,\mu_l)&=\beta_lx - \delta_l(x-K) +\lambda_l  x \alpha_1(x) + \lambda_l x^{\gamma^+_h} \alpha_2(x)-\lambda_l \delta_h K \alpha_3(x)
\end{aligned}
\end{align}
with
\begin{align}
\begin{aligned} \label{defvaralpha}
    &\alpha_1 (x) = \int_0^\infty \e^{-(r_l +\alpha-\theta)t}[\beta_h\Phi(-d_1(x,t))+\delta_h\Phi(d_1(x,t))]  \ud t \\
    &\alpha_2 (x) = \int_0^\infty \e^{(\Delta -\lambda_l)t} \Phi(-d_2(x,t)-\sigma\gamma^+_h\sqrt{t})  \ud t \\
    &\alpha_3(x) = \int_0^\infty \e^{-r_l t} \Phi(d_2(x,t))  \ud t .
\end{aligned}
\end{align}

Our goal is to compute the coefficients $\alpha_1$, $\alpha_2$, and $\alpha_3$ explicitly. We detail the computation of $\alpha_1$ only, as the procedures for $\alpha_2$ and $\alpha_3$ follow analogously with minor adjustments. To simplify the exposition, let $r\coloneqq(r_l +\alpha-\theta)$.
The computation of $\alpha_1$ is structured into four steps for clarity.
\begin{itemize}
    \item[\textit{Step 1.}]  We show that 
    \begin{align}
        \begin{aligned} \label{Step1}
                r\int_0^\infty \e^{-rt}\Phi(d_1(x,t))  \ud t  &=  \ind_{\{x> x^4_h\}}+\frac{1}{2}\ind_{\{x=x^4_h\}}\\
                &\ \ +\frac{1}{2} \int_0^\infty \frac{\e^{-rt} \e^{-\frac{d_1(x,t)^2}{2}} }{\sqrt{2\pi}\sigma \sqrt{t}} \frac{\big(\theta-\alpha+\frac{\sigma^2}{2}\big)t-\ln\big(\frac{x}{x^4_h}\big)}{t}  \ud t .
        \end{aligned}
    \end{align}
    We start by observing that by partial integration we have \begin{align}
        \begin{aligned}
                r\int_0^\infty \e^{-rt}\Phi(d_1(x,t))  \ud t  &= \bigg[-\e^{-rt}\int_{-\infty}^{d_1(x,t)} \frac{\e^{-\frac{z^2}{2}}}{\sqrt{2\pi}} dz\bigg]_0^\infty \\
                &\ \ +\frac{1}{2} \int_0^\infty \frac{\e^{-rt} \e^{-\frac{d_1(x,t)^2}{2}} }{\sqrt{2\pi}\sigma \sqrt{t}} \frac{\big(\theta-\alpha+\frac{\sigma^2}{2}\big)t-\ln\big(\frac{x}{x^4_h}\big)}{t}  \ud t .
        \end{aligned}
    \end{align}
    Then, since  
    \begin{equation}
    \lim_{t\to\infty} -\e^{-rt}\int_{-\infty}^{d_1(x,t)} \frac{\e^{-\frac{z^2}{2}}}{\sqrt{2\pi}} dz=0 \end{equation}
    and 
    \begin{equation} \lim_{t\to0} d_1(x,t)=\begin{cases}
    \begin{alignedat}{2}
        &-\infty, &&\ \ \ \text{if} \ \ x<x^4_h,\\
        &0, &&\ \ \ \text{if} \ \ x=x^4_h,\\
        &+\infty, &&\ \ \ \text{if} \ \ x>x^4_h,
        \end{alignedat}
    \end{cases} \end{equation}
     we get
    \begin{align}
        \begin{aligned}
                \bigg[-\e^{-rt}\int_{-\infty}^{d_1(x,t)} \frac{\e^{-\frac{z^2}{2}}}{\sqrt{2\pi}} dz\bigg]_0^\infty = \ind_{\{x> x^4_h\}}+\frac{1}{2}\ind_{\{x=x^4_h\}}.
        \end{aligned}
    \end{align}
    This concludes \textit{Step 1}.
    
    \item[\textit{Step 2.}] We now show that \begin{align}
        \begin{aligned} \label{step2eq}
                &r\int_0^\infty \e^{-rt}\Phi(d_1(x,t))  \ud t  - \frac{ a_1 }{2} \int_0^\infty \frac{\e^{-rt} \e^{-\frac{d_1(x,t)^2}{2}} }{\sqrt{2\pi}\sigma \sqrt{t}}  \ud t  =\\
                &\quad =\ind_{\{x> x^4_h\}} + \bigg( \frac{x^4_h}{x} \bigg)^{\frac{ a_1 }{\sigma^2}} \ind_{\{x< x^4_h\}}+\frac{1}{2}\bigg[ 1+ \bigg( \frac{x^4_h}{x} \bigg)^{\frac{ a_1 }{\sigma^2}} \bigg]\ind_{\{x=x^4_h\}}
        \end{aligned}
    \end{align}
    with
    \begin{equation} \label{a1}  a_1 \coloneqq\theta-\alpha+\frac{\sigma^2}{2}-\sqrt{\bigg(\theta-\alpha+\frac{\sigma^2}{2}\bigg)^2+2r\sigma^2}.  \end{equation}
    
    Using \eqref{Step1} from \textit{Step 1}, it is sufficient to show that 
    \begin{align}
        \begin{aligned} \label{Step22}
                &\frac{1}{2} \int_0^\infty \frac{\e^{-rt} \e^{-\frac{d_1(x,t)^2}{2}} }{\sqrt{2\pi}\sigma \sqrt{t}} \bigg[\frac{\big(\theta-\alpha+\frac{\sigma^2}{2}\big)t-\ln\big(\frac{x}{x^4_h}\big)}{t} - a_1 \bigg]  \ud t  = \\
                &\quad \bigg( \frac{x^4_h}{x} \bigg)^{\frac{ a_1 }{\sigma^2}} \ind_{\{x< x^4_h\}}+\frac{1}{2}\bigg( \frac{x^4_h}{x} \bigg)^{\frac{ a_1 }{\sigma^2}} \ind_{\{x=x^4_h\}}.
        \end{aligned}
    \end{align}

    To this end, we define \begin{equation} \label{g1} g_1(t)\coloneqq \frac{\sqrt{\big(\theta-\alpha+\frac{\sigma^2}{2}\big)^2+2r\sigma^2} \ t+\ln\big(\frac{x}{x^4_h}\big)}{\sigma \sqrt{t}}. \end{equation}
    We now compute \begin{align}
        \begin{aligned} \label{exponentistep}
                &d_1(x,t)^2+2rt =\\
                &\quad =\frac{2\ln\big(\frac{x}{x^4_h}\big) + \big(\theta-\alpha+\frac{\sigma^2}{2}\big)^2 t^2 + 2 \big(\theta-\alpha+\frac{\sigma^2}{2}\big)t \ln\big(\frac{x}{x^4_h}\big) + 2r \sigma^2 t^2}{\sigma^2 t}\\
                &\quad =\bigg[ \frac{\sqrt{\big(\theta-\alpha+\frac{\sigma^2}{2}\big)^2+2r\sigma^2} \ t+\ln\big(\frac{x}{x^4_h}\big)}{\sigma \sqrt{t}}  \bigg]^2 \\
                &\quad \quad + 2 \frac{\big(\theta-\alpha+\frac{\sigma^2}{2}\big)-\sqrt{\big(\theta-\alpha+\frac{\sigma^2}{2}\big)^2+2r\sigma^2}}{\sigma^2}\ln\bigg(\frac{x}{x^4_h}\bigg)\\
                &\quad =g_1(t)^2+2\frac{ a_1 }{\sigma^2}\ln\bigg(\frac{x}{x^4_h}\bigg).
        \end{aligned}
    \end{align}
    Moreover, from the definition in \eqref{g1}, it follows that 
    \begin{equation} \label{g1prime} g_1'(t)= \frac{\sqrt{\big(\theta-\alpha+\frac{\sigma^2}{2}\big)^2+2r\sigma^2} \ t - \ln\big(\frac{x}{x^4_h}\big)}{2t\sigma \sqrt{t}}, \end{equation}
    and
    \begin{equation} \label{glimit} \lim_{t\to+\infty} g_1(t)=+\infty \qquad \text{and} \qquad \lim_{t\to 0} g_1(t)=\begin{cases}
    \begin{alignedat}{2}
    &-\infty, &\ \ \ \text{if} \ \ x<x^4_h,\\
    &0, &\ \ \ \text{if} \ \ x=x^4_h,\\
    &+\infty, &\ \ \ \text{if} \ \ x>x^4_h.  
    \end{alignedat}  
    \end{cases} \end{equation}
    Finally we prove \eqref{Step22}. We have 
    \begin{align}
        \begin{aligned} \label{step2final}
                &\frac{1}{2} \int_0^\infty \frac{\e^{-rt} \e^{-\frac{d_1(x,t)^2}{2}} }{\sqrt{2\pi}\sigma \sqrt{t}} \bigg[\frac{\big(\theta-\alpha+\frac{\sigma^2}{2}\big)t-\ln\big(\frac{x}{x^4_h}\big)}{t} - a_1 \bigg]  \ud t  \\
                &=\frac{1}{2} \int_0^\infty \frac{\e^{-rt} \e^{-\frac{d_1(x,t)^2}{2}} }{\sqrt{2\pi}\sigma \sqrt{t}} \frac{\sqrt{\big(\theta-\alpha+\frac{\sigma^2}{2}\big)^2+2r\sigma^2} \ t-\ln\big(\frac{x}{x^4_h}\big)}{t}  \ud t  =\\
                &=\bigg(\frac{x^4_h}{x}\bigg)^{\frac{ a_1 }{\sigma^2}} \int_0^\infty \frac{\e^{-\frac{g_1(t)^2}{2}}}{\sqrt{2\pi}} g_1'(t)  \ud t =\bigg(\frac{x^4_h}{x}\bigg)^{\frac{ a_1 }{\sigma^2}} \int_{g_1(0)}^{g_1(\infty)} \frac{\e^{-\frac{q^2}{2}}}{\sqrt{2\pi}}dq\\
                &=\bigg( \frac{x^4_h}{x} \bigg)^{\frac{ a_1 }{\sigma^2}} \ind_{\{x< x^4_h\}} +\frac{1}{2} \bigg( \frac{x^4_h}{x} \bigg)^{\frac{ a_1 }{\sigma^2}} \ind_{\{x=x^4_h\}},
        \end{aligned}
    \end{align}   
    where in the first equality we used the definition of $a_1$ in \eqref{a1}, the second equality applies \eqref{exponentistep}, \eqref{g1}, and \eqref{g1prime}. The third equality follows from a change of variable in the integral and the last equality uses the limit behavior of $g$ in \eqref{glimit}.

    \item[\textit{Step 3.}] We now show that \begin{align}
        \begin{aligned} \label{step3eq}
                &r\int_0^\infty \e^{-rt}\Phi(d_1(x,t))  \ud t  - \frac{ a_2 }{2} \int_0^\infty \frac{\e^{-rt} \e^{-\frac{d_1(x,t)^2}{2}} }{\sqrt{2\pi}\sigma \sqrt{t}}  \ud t  =\\
                &\quad \ind_{\{x> x^4_h\}} + \bigg( \frac{x^4_h}{x} \bigg)^{\frac{ a_2 }{\sigma^2}} \ind_{\{x< x^4_h\}}+\frac{1}{2}\bigg[ 1+ \bigg( \frac{x^4_h}{x} \bigg)^{\frac{ a_2 }{\sigma^2}} \bigg]\ind_{\{x=x^4_h\}}
        \end{aligned}
    \end{align}
    with
    \begin{equation}  a_2 \coloneqq\theta-\alpha+\frac{\sigma^2}{2}+\sqrt{\bigg(\theta-\alpha+\frac{\sigma^2}{2}\bigg)^2+2r\sigma^2}.  \end{equation}
The proof closely mirrors that of \textit{Step 2}. Using \eqref{Step1} from \textit{Step 1}, it is sufficient to prove that
    \begin{align}
        \begin{aligned} \label{step33}
                &\frac{1}{2} \int_0^\infty \frac{\e^{-rt} \e^{-\frac{d_1(x,t)^2}{2}} }{\sqrt{2\pi}\sigma \sqrt{t}} \bigg[\frac{\big(\theta-\alpha+\frac{\sigma^2}{2}\big)t-\ln\big(\frac{x}{x^4_h}\big)}{t} - a_2 \bigg]  \ud t  =\\
                &\quad =\bigg( \frac{x^4_h}{x} \bigg)^{\frac{ a_2 }{\sigma^2}} \ind_{\{x< x^4_h\}}+\frac{1}{2}\bigg( \frac{x^4_h}{x} \bigg)^{\frac{ a_2 }{\sigma^2}} \ind_{\{x=x^4_h\}}.
        \end{aligned}
    \end{align}
    
    Letting \begin{equation} g_2(t)\coloneqq \frac{\ln\big(\frac{x}{x^4_h}\big)-\sqrt{\big(\theta-\alpha+\frac{\sigma^2}{2}\big)^2+2r\sigma^2} \ t}{\sigma \sqrt{t}}, \end{equation}
    and proceeding as in \eqref{exponentistep}, we get
    \begin{align}
        \begin{aligned}
                d_1(x,t)^2+2rt &=\frac{2\ln\big(\frac{x}{x^4_h}\big) + \big(\theta-\alpha+\frac{\sigma^2}{2}\big)^2 t^2 + 2 \big(\theta-\alpha+\frac{\sigma^2}{2}\big)t \ln\big(\frac{x}{x^4_h}\big) + 2r \sigma^2 t^2}{\sigma^2 t}\\
                &\quad =\bigg[ \frac{\ln\big(\frac{x}{x^4_h}\big)-\sqrt{\big(\theta-\alpha+\frac{\sigma^2}{2}\big)^2+2r\sigma^2} \ t}{\sigma \sqrt{t}}  \bigg]^2 \\
                &\quad \quad  + 2 \frac{\big(\theta-\alpha+\frac{\sigma^2}{2}\big)+\sqrt{\big(\theta-\alpha+\frac{\sigma^2}{2}\big)^2+2r\sigma^2}}{\sigma^2}\ln\bigg(\frac{x}{x^4_h}\bigg)\\
                &\quad =
                g_2(t)^2+2\frac{ a_2 }{\sigma^2}\ln\bigg(\frac{x}{x^4_h}\bigg).
        \end{aligned}
    \end{align}

    Observing that  
    \begin{equation} g_2'(t)= -\frac{\sqrt{\big(\theta-\alpha+\frac{\sigma^2}{2}\big)^2+2r\sigma^2} \ t + \ln\big(\frac{x}{x^4_h}\big)}{2t\sigma \sqrt{t}} \end{equation}
    and
    \begin{equation} \lim_{t\to+\infty} g_2(t)=+\infty \qquad \text{and} \qquad \lim_{t\to 0} g_2(t)=\begin{cases}
    \begin{alignedat}{2}
    &-\infty, &\ \ \ \text{if} \ \ x<x^4_h,\\
    &0, &\ \ \ \text{if} \ \ x=x^4_h,\\
    &+\infty, &\ \ \ \text{if} \ \ x>x^4_h,
    \end{alignedat}
    \end{cases} \end{equation}
    we can proceed as in the proof of \eqref{step2final} to finally prove \eqref{step33}.
    Specifically,
    \begin{align}
        \begin{aligned}
            &\frac{1}{2} \int_0^\infty \frac{\e^{-rt} \e^{-\frac{d_1(x,t)^2}{2}} }{\sqrt{2\pi}\sigma \sqrt{t}} \bigg[\frac{\big(\theta-\alpha+\frac{\sigma^2}{2}\big)t-\ln\big(\frac{x}{x^4_h}\big)}{t} - a_2 \bigg]  \ud t  =\\
            &= -\frac{1}{2} \int_0^\infty \frac{\e^{-rt} \e^{-\frac{d_1(x,t)^2}{2}} }{\sqrt{2\pi}\sigma \sqrt{t}} \frac{\sqrt{\big(\theta-\alpha+\frac{\sigma^2}{2}\big)^2+2r\sigma^2} \ t+\ln\big(\frac{x}{x^4_h}\big)}{t}  \ud t  =\\
            &=\bigg(\frac{x^4_h}{x}\bigg)^{\frac{ a_2 }{\sigma^2}} \int_0^\infty \frac{\e^{-\frac{g_2(t)^2}{2}}}{\sqrt{2\pi}} g_2'(t)  \ud t  =\bigg(\frac{x^4_h}{x}\bigg)^{\frac{ a_2 }{\sigma^2}} \int_{g_2(0)}^{g_2(\infty)} \frac{\e^{-\frac{q^2}{2}}}{\sqrt{2\pi}}dq\\
            &=\bigg( \frac{x^4_h}{x} \bigg)^{\frac{ a_2 }{\sigma^2}} \ind_{\{x< x^4_h\}} +\frac{1}{2} \bigg( \frac{x^4_h}{x} \bigg)^{\frac{ a_2 }{\sigma^2}} \ind_{\{x=x^4_h\}},
        \end{aligned}
    \end{align} 
    that concludes the Step 3.

\item[\textit{Step 4.}] We finally derive the expression of $\alpha_1$ as defined in Proposition 3.5 of the main text. 
We first observe that from the system 
    \begin{equation} 
    \begin{cases}
        r z_1 - \frac{ a_1 }{2} z_2 = \ind_{\{x> x^4_h\}} + \Big( \frac{x^4_h}{x} \Big)^{\frac{ a_1 }{\sigma^2}} \ind_{\{x< x^4_h\}}+\frac{1}{2}\Big[ 1+ \Big( \frac{x^4_h}{x} \Big)^{\frac{ a_1 }{\sigma^2}} \Big]\ind_{\{x=x^4_h\}} \\
        r z_1 - \frac{ a_2 }{2} z_2 = \ind_{\{x> x^4_h\}} + \Big( \frac{x^4_h}{x} \Big)^{\frac{ a_2 }{\sigma^2}} \ind_{\{x< x^4_h\}}+\frac{1}{2}\Big[ 1+ \Big( \frac{x^4_h}{x} \Big)^{\frac{ a_2 }{\sigma^2}} \Big]\ind_{\{x=x^4_h\}},
    \end{cases} \end{equation}
we find
        \begin{align}
        \begin{aligned} \label{zeta1step4}
                z_1&= \frac{1}{r}( \ind_{\{x> x^4_h\}}+\frac{1}{2}\ind_{\{x=x^4_h\}}) \\
                &\quad+\frac{ a_1 }{r( a_1 - a_2 )} \Big[ \Big( \frac{x^4_h}{x} \Big)^{\frac{ a_2 }{\sigma^2}} -\frac{ a_2 }{ a_1 } \Big( \frac{x^4_h}{x} \Big)^{\frac{ a_1 }{\sigma^2}}\Big] \big( \ind_{\{x< x^4_h\}}+\frac{1}{2}\ind_{\{x=x^4_h\}}\big).
        \end{aligned}
    \end{align}
    Then, combining \eqref{step2eq} from \textit{Step 2} and \eqref{step3eq} from \textit{Step 3}, we find from \eqref{zeta1step4} that 
    \begin{align}
        \begin{aligned} \label{zeta1step42}
                &\int_0^\infty \e^{-rt}\Phi(d_1(x,t))  \ud t = \frac{1}{r}( \ind_{\{x> x^4_h\}}+\frac{1}{2}\ind_{\{x=x^4_h\}})\\
                &\ \ \ +\frac{ a_1 }{r( a_1 - a_2 )} \Big[ \Big( \frac{x^4_h}{x} \Big)^{\frac{ a_2 }{\sigma^2}} -\frac{ a_2 }{ a_1 } \Big( \frac{x^4_h}{x} \Big)^{\frac{ a_1 }{\sigma^2}}\Big] \big( \ind_{\{x< x^4_h\}}+\frac{1}{2}\ind_{\{x=x^4_h\}}\big).
        \end{aligned}
    \end{align}

    Notice that 
    \begin{align}
        \begin{aligned}
                a_1&=\frac{\theta-\alpha}{\sigma^2}+\frac{1}{2}- \sqrt{\bigg(\frac{\theta-\alpha}{\sigma^2}-\frac{1}{2}\bigg)^2+\frac{2r_l }{\sigma^2}}=\sigma^2(1-\gamma^+_l),
        \end{aligned}
    \end{align}
    and similarly $a_2=\sigma^2(1-\gamma^-_l)$. Thus, from \eqref{zeta1step42}, we get
    \begin{align}
        \begin{aligned} \label{step4eq}
                \int_0^\infty \e^{-rt}\Phi(d_1(x,t))  \ud t  
                &= \frac{1}{r} \ind_{\{x\geq x^4_h\}}+\frac{1-\gamma^+_l}{r(\gamma^-_l-\gamma^+_l)} \bigg[ \bigg( \frac{x^4_h}{x} \bigg)^{1-\gamma^-_l} \\
                &-\frac{1-\gamma^-_l}{1-\gamma^+_l} \bigg( \frac{x^4_h}{x} \bigg)^{1-\gamma^+_l}\bigg]  \ind_{\{x< x^4_h\}}.
        \end{aligned}
    \end{align}

Finally, from \eqref{defvaralpha}, we get 
    \begin{align} \label{alpha1eq}
        \begin{aligned}
                \alpha_1 (x) &= \int_0^\infty \e^{-r t}[\beta_h+(\delta_h-\beta_h)\Phi(d_1(x,t))]  \ud t \\
                &= \frac{\beta_h}{r} + (\delta_h-\beta_h)\int_0^\infty \e^{-r t}\Phi(d_1(x,t))  \ud t ,
        \end{aligned}
    \end{align}
and combining the latter expression with \eqref{step4eq} we derive $\alpha_1$ as defined in Proposition 3.5 of the main text. 
\end{itemize}
To compute $\alpha_2$ and $\alpha_3$, one can follow the same procedure as in the derivation of $\alpha_1$. 
Indeed, letting \begin{equation} 
a_{3,4}\coloneqq\theta-\alpha +\sigma^2\bigg(\gamma^+_h-\frac{1}{2}\bigg)\mp\sqrt{\bigg(\theta-\alpha+\sigma^2\bigg(\gamma^+_h-\frac{1}{2}\bigg)\bigg)^2+2\sigma^2(\lambda_l-\Delta)},\end{equation} 
and following \textit{Step 1--5} above, one can show that
\begin{align}
    \begin{aligned} 
        \alpha_2 = \frac{1}{\lambda_l-\Delta } \ind_{\{x\geq x^4_h\}} + \frac{ a_3 }{(\lambda_l-\Delta )( a_3 - a_4 )} \bigg[ \bigg( \frac{x^4_h}{x} \bigg)^{\frac{ a_4 }{\sigma^2}}-\frac{ a_4 }{ a_3 } \bigg( \frac{x^4_h}{x} \bigg)^{\frac{ a_3 }{\sigma^2}}\bigg]  \ind_{\{x< x^4_h\}}.
    \end{aligned}
\end{align}
Then, since   \begin{align}
    \begin{aligned}
        \frac{ a_3 }{\sigma^2}
        &=\sqrt{\bigg( \frac{1}{2} -\frac{\theta-\alpha}{\sigma^2}\bigg)^2 + \frac{2r_h}{\sigma^2}}-\sqrt{\bigg( \frac{1}{2} -\frac{\theta-\alpha}{\sigma^2}\bigg)^2 + \frac{2r_l }{\sigma^2}}=\gamma^+_h- \gamma^+_l,
    \end{aligned}
\end{align}
and  $a_4=\sigma(\gamma^+_h-\gamma^-_l)$. One easily finds the explicit expression for $\alpha_2$ as defined in Proposition 3.5 of the main text. 
Similarly, one can also show that 
\begin{align}
    \begin{aligned} 
        \alpha_3(x) &= \int_0^\infty \e^{-r_l  t} \Phi(d_2(x,t))  \ud t \\
        &= \frac{1}{r_l }\ind_{\{x\geq x^4_h\}} -\frac{ \gamma^+_l}{r_l (\gamma^-_l-\gamma^+_l)} \bigg[ \bigg( \frac{x^4_h}{x} \bigg)^{- \gamma^-_l} -\frac{ \gamma^-_l}{\gamma^+_l} \bigg( \frac{x^4_h}{x} \bigg)^{-\gamma^+_l}\bigg] \ind_{\{x< x^4_h\}}.
    \end{aligned}
\end{align}

Then, combining the resulting expression for $\alpha_1$, $\alpha_2$ and $\alpha_3$ with \eqref{defvaralpha}, we derive the explicit form of $W(x,\mu_l)$. Recalling the relation \eqref{ValuefunctionWonejump} between $V$ and $W$, the proof is complete.

\end{proof}

\begin{proof}[\textbf{Proof of Proposition 3.6 in the main text}]
From Proposition \ref{proposuffcondm}-(3), if $M_l \le 0$ then $\cC_{l}=\varnothing$ and $\cS_l=[0,\infty)$. In this case, clearly $W(x,\mu_l)\equiv 0$ and, using \eqref{ValuefunctionWonejump}, $V(x,\mu_l) = \delta_l (x-K)$ for $x\ge 0$. If $M_l>0$ then, from Proposition \ref{proposuffcondm}-(3), it follows that $\cC_{l}=(0,\infty)$ and $\cS_l=\{0\}$. In this case, $M(x,\mu_l)= M_l x$ and
\begin{align}
    W(x,\mu_l)&= \E_x \bigg[\int_0^\infty e^{-r_l t} M(X_t,\mu_l) dt \bigg] = \int_0^\infty e^{-r_l t} M_l \E_x[X_t] dt = \frac{M_l}{r_l + \alpha - \theta} x.
\end{align}
From \eqref{ValuefunctionWonejump}, the proposition follows.

\end{proof}

\bibliographystyle{plain}
\bibliography{biblio}